\documentclass[twoside,twocolumn,9pt]{article}
\usepackage{extsizes}
\usepackage[super,sort&compress,comma]{natbib} 
\usepackage[version=3]{mhchem}
\usepackage[left=1.5cm, right=1.5cm, top=1.785cm, bottom=2.0cm]{geometry}
\usepackage{balance}
\usepackage{mathptmx}
\usepackage{sectsty}
\usepackage{graphicx} 
\usepackage{lastpage}
\usepackage[format=plain,justification=justified,singlelinecheck=false,font={stretch=1.125,small,sf},labelfont=bf,labelsep=space]{caption}
\usepackage{float}
\usepackage{fancyhdr}
\usepackage{fnpos}
\usepackage[english]{babel}
\addto{\captionsenglish}{%
  
}
\usepackage{array}
\usepackage{droidsans}
\usepackage{charter}
\usepackage[T1]{fontenc}
\usepackage[usenames,dvipsnames]{xcolor}
\usepackage{setspace}
\usepackage[compact]{titlesec}
\usepackage{hyperref}

\usepackage{epstopdf}

\usepackage{booktabs}
\usepackage[referable]{threeparttablex}
\usepackage{multirow}
\usepackage{amsmath, amssymb}

\definecolor{cream}{RGB}{222,217,201}

\DeclareMathOperator*{\argmin}{arg\,min}
\DeclareMathOperator{\EX}{\mathbb{E}}

\begin{document}

\pagestyle{fancy}
\thispagestyle{plain}
\fancypagestyle{plain}{
\renewcommand{\headrulewidth}{0pt}
}

\makeFNbottom
\makeatletter
\renewcommand\LARGE{\@setfontsize\LARGE{15pt}{17}}
\renewcommand\Large{\@setfontsize\Large{12pt}{14}}
\renewcommand\large{\@setfontsize\large{10pt}{12}}
\renewcommand\footnotesize{\@setfontsize\footnotesize{7pt}{10}}
\makeatother

\renewcommand{\thefootnote}{\fnsymbol{footnote}}
\renewcommand\footnoterule{\vspace*{1pt}%
\color{cream}\hrule width 3.5in height 0.4pt \color{black}\vspace*{5pt}} 
\setcounter{secnumdepth}{5}

\makeatletter 
\renewcommand\@biblabel[1]{#1}            
\renewcommand\@makefntext[1]%
{\noindent\makebox[0pt][r]{\@thefnmark\,}#1}
\makeatother 
\renewcommand{\figurename}{\small{Fig.}~}
\sectionfont{\sffamily\Large}
\subsectionfont{\normalsize}
\subsubsectionfont{\bf}
\setstretch{1.125} 
\setlength{\skip\footins}{0.8cm}
\setlength{\footnotesep}{0.25cm}
\setlength{\jot}{10pt}
\titlespacing*{\section}{0pt}{4pt}{4pt}
\titlespacing*{\subsection}{0pt}{15pt}{1pt}

\fancyfoot{}
\fancyfoot[LO,RE]{\vspace{-7.1pt}\includegraphics[height=9pt]{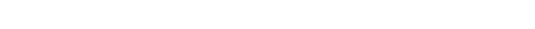}}
\fancyfoot[CO]{\vspace{-7.1pt}\hspace{13.2cm}\includegraphics{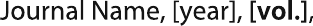}}
\fancyfoot[CE]{\vspace{-7.2pt}\hspace{-14.2cm}\includegraphics{head_foot/RF}}
\fancyfoot[RO]{\footnotesize{\sffamily{1--\pageref{LastPage} ~\textbar  \hspace{2pt}\thepage}}}
\fancyfoot[LE]{\footnotesize{\sffamily{\thepage~\textbar\hspace{3.45cm} 1--\pageref{LastPage}}}}
\fancyhead{}
\renewcommand{\headrulewidth}{0pt} 
\renewcommand{\footrulewidth}{0pt}
\setlength{\arrayrulewidth}{1pt}
\setlength{\columnsep}{6.5mm}
\setlength\bibsep{1pt}

\makeatletter 
\newlength{\figrulesep} 
\setlength{\figrulesep}{0.5\textfloatsep} 

\newcommand{\topfigrule}{\vspace*{-1pt}%
\noindent{\color{cream}\rule[-\figrulesep]{\columnwidth}{1.5pt}} }

\newcommand{\botfigrule}{\vspace*{-2pt}%
\noindent{\color{cream}\rule[\figrulesep]{\columnwidth}{1.5pt}} }

\newcommand{\dblfigrule}{\vspace*{-1pt}%
\noindent{\color{cream}\rule[-\figrulesep]{\textwidth}{1.5pt}} }

\makeatother

\twocolumn[
  \begin{@twocolumnfalse}
{\includegraphics[height=30pt]{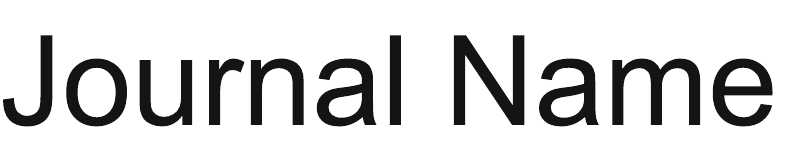}\hfill\raisebox{0pt}[0pt][0pt]{\includegraphics[height=55pt]{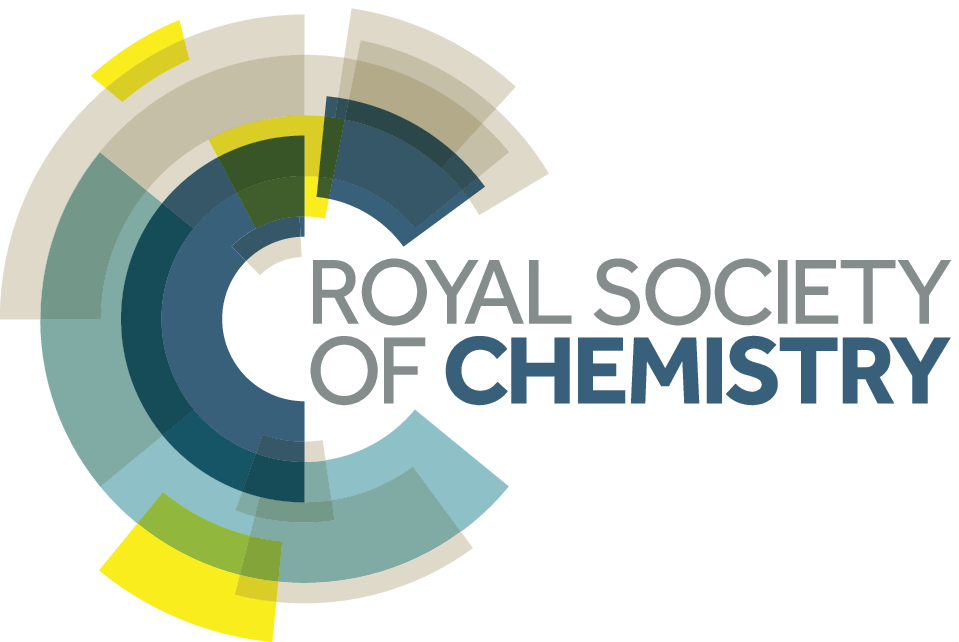}}\\[1ex]
\includegraphics[width=18.5cm]{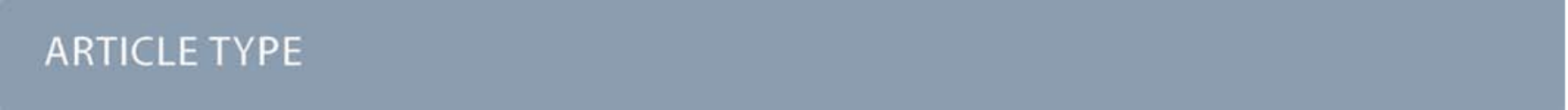}}\par
\vspace{1em}
\sffamily
\begin{tabular}{m{4.5cm} p{13.5cm} }

\includegraphics{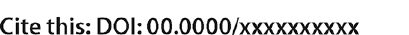} & \noindent\LARGE{\textbf{Materials Discovery with Extreme Properties via Reinforcement Learning-Guided Combinatorial Chemistry$^\dag$}} \\
\vspace{0.3cm} & \vspace{0.3cm} \\

 & \noindent\large{Hyunseung Kim,\textit{$^{a,\ddag}$} Haeyeon Choi,\textit{$^{b,c,\ddag}$} Dongju Kang,\textit{$^{a,\ddag}$}} Won Bo Lee,$^{\ast}$\textit{$^{a}$} and Jonggeol Na$^{\ast}$\textit{$^{b,c}$}\\

\includegraphics{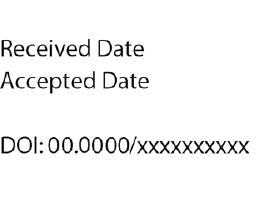} & \noindent\normalsize{The goal of most materials discovery is to discover materials that are superior to those currently known. Fundamentally, this is close to extrapolation, which is a weak point for most machine learning models that learn the probability distribution of data. Herein, we develop reinforcement learning-guided combinatorial chemistry, which is a rule-based molecular designer driven by trained policy for selecting subsequent molecular fragments to get a target molecule. Since our model has the potential to generate all possible molecular structures that can be obtained from combinations of molecular fragments, unknown molecules with superior properties can be discovered. We theoretically and empirically demonstrate that our model is more suitable for discovering better compounds than probability distribution-learning models. In an experiment aimed at discovering molecules that hit seven extreme target properties, our model discovered 1,315 of all target-hitting molecules and 7,629 of five target-hitting molecules out of 100,000 trials, whereas the probability distribution-learning models failed. Moreover, it has been confirmed that every molecule generated under the binding rules of molecular fragments is 100\% chemically valid. To illustrate the performance in actual problems, we also demonstrate that our models work well on two practical applications: discovering protein docking molecules and HIV inhibitors.} \\

\end{tabular}

 \end{@twocolumnfalse} \vspace{0.6cm}

  ]

\renewcommand*\rmdefault{bch}\normalfont\upshape
\rmfamily
\section*{}
\vspace{-1cm}


\footnotetext{\textit{$^{a}$~School of Chemical and Biological Engineering, Seoul National University, Republic of Korea; E-mail: wblee@snu.ac.kr}}
\footnotetext{\textit{$^{b}$~Department of Chemical Engineering and Materials Science, Ewha Womans University, Republic of Korea; E-mail: jgna@ewha.ac.kr}}
\footnotetext{\textit{$^{c}$~Graduate Program in System Health Science and Engineering, Ewha Womans University, Republic of Korea}}

\footnotetext{\dag~Electronic Supplementary Information (ESI) available: [details of any supplementary information available should be included here]. See DOI: 00.0000/00000000.}
\footnotetext{\textit{$^{\ddag}$~These authors contributed equally to this work.}}




\section{Introduction}
The task of discovering materials that are superior to those currently known is a challenging problem in various fields of materials science, including pharmaceutical substances \cite{pommier2005integrase, yang2021hit, kitchen2004docking, gottipati2020learning}, electrical and electronic materials \cite{klein2010conductors, greenaway2021integrating, sylvinson2019OLED, kim2018OLED, LOCO, ext_limit1, pyzer2015HTS}, energy materials \cite{pyzer2015HTS, liu2022QSPR, dan2020unconditionedGAN, sv2022multi}, metals and ceramics \cite{ext_limit1}, nanomaterials \cite{dong2020RCGAN}, and polymeric materials \cite{lyu2015plastics, stauber2007plastics}. Some of these studies aim to discover materials with two or more target properties that contradict each other, meaning it is difficult for them to coexist \cite{zunger2018inverse}. For example, super engineering plastics used in automobiles should be lighter than metals yet have similar mechanical strength \cite{lyu2015plastics, stauber2007plastics}. Similarly, transparent conductors used in display panels should be both optically transparent (requiring a large bandgap) and electrically conductive (requiring high carrier concentration, which generally has a low bandgap) \cite{zunger2018inverse, klein2010conductors}. In some cases, the aim is to discover materials that have properties with either extremely high or low values. For example, the development of a better organic light-emitting diode (OLED) requires chemists to discover novel materials with higher efficiency and stability \cite{greenaway2021integrating, sylvinson2019OLED, kim2018OLED}. Here, the problem is that there are no (or few) known samples that have such properties compared to common substances. This makes it difficult for chemists to gain insights or knowledge from the known materials, that could help to infer the molecular structures of the desired materials. Unfortunately, this situation also holds for most machine learning models that learn the data. Therefore, it is necessary to develop a model that can discover materials, even in regions with little or no known data. In this paper, we refer to this problem as \textit{materials extrapolation}.

\begin{figure*}[h!]
    \centering
    \includegraphics[width=0.95\textwidth]{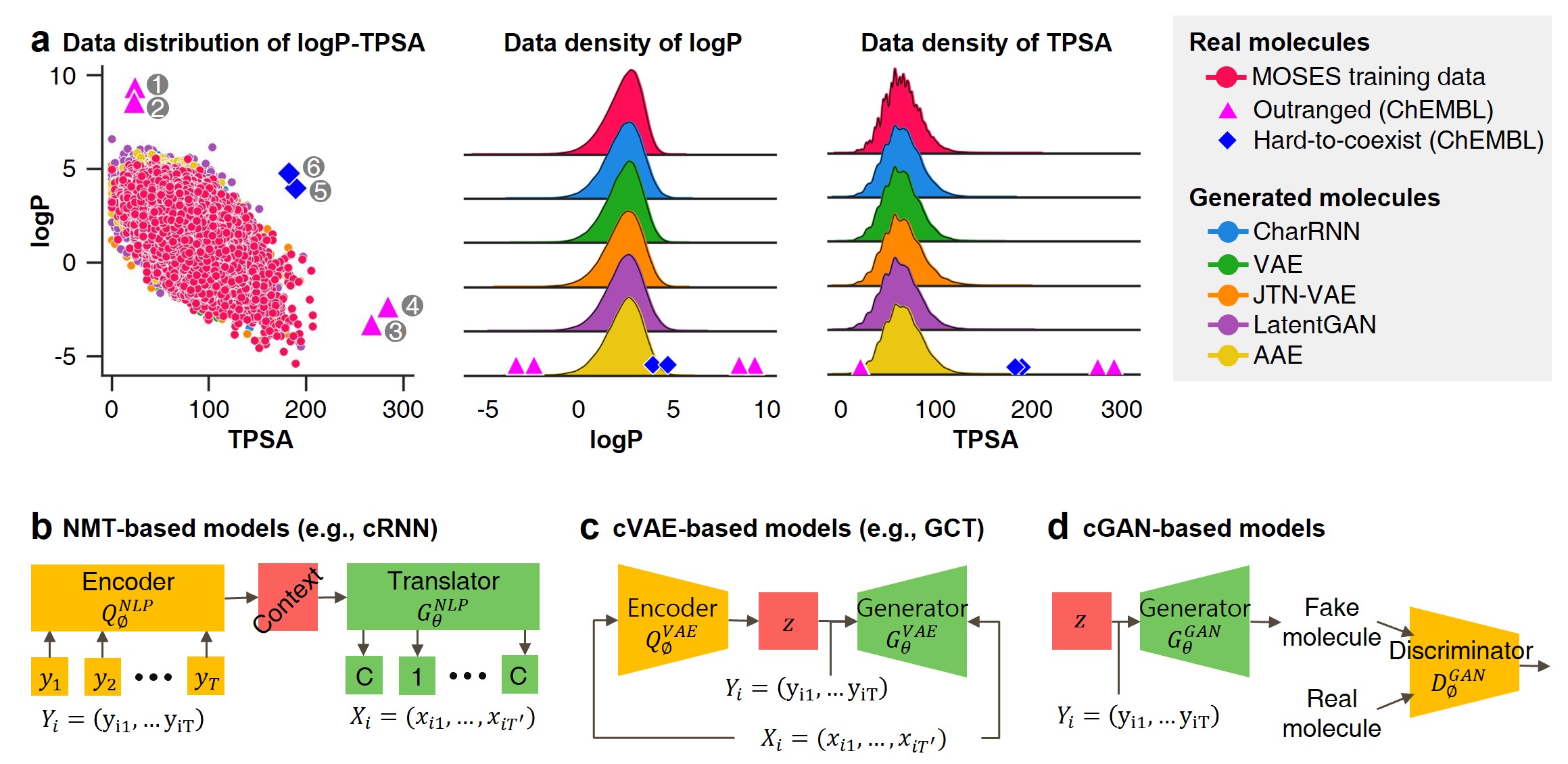}
    \caption{\textbf{Probability distribution-learning models for molecular generation.} (a) Data distribution of logP-TPSA. The pink dots denote the molecules in MOSES \cite{polykovskiy2020moses} training data. The other colored dots denote the molecules generated by MOSES baseline models which were trained with the MOSES training data. Since the MOSES baseline models are probability distribution-learning models such as NMT, GAN, VAE, and AAE, the distribution of generated molecules approximates the distribution of their training data. The magenta triangles and blue diamonds indicate real molecules in ChEMBL \cite{mendez2019chembl} database, which have extrapolated properties from MOSES training data distribution. \textcircled{1} CHEMBL3216345;
 \textcircled{2} CHEMBL3230084; \textcircled{3} CHEMBL3358630; \textcircled{4} CHEMBL300801; \textcircled{5} CHEMBL501130; \textcircled{6} CHEMBL52004. (b-d) Types of inverse molecular designer. $X_i$, $Y_i$, and $z$ denote $i$-th molecular structure, properties of the $i$-th molecule, and latent code, respectively.}\label{fig:moses_distribution}
\end{figure*}

In recent years, it has been reported that machine learning techniques can solve many challenging problems in a wide range of fields, including materials discovery. In particular, models for goal-directed inverse molecular design are attractive because they can directly infer the molecular structures that meet a set of given target conditions such as scaffolds \cite{mendez2020cGAN, yang2021hit}, physical properties \cite{dong2020RCGAN, lim2018cVAE_seongokRyu, Kim2021GCT, kotsias2020cRNN}, and biological activities \cite{kotsias2020cRNN, gottipati2020learning, yang2021hit}. Some of these studies have proposed models based on neural machine translation (NMT) such as seq2seq \cite{sutskever2014seq2seq_original, kotsias2020cRNN} and Transformer \cite{vaswani2017Transformer_original}, which translate input target conditions to corresponding molecular structures. Models based on conditional generative models have also been proposed, such as conditional generative adversarial networks (cGANs) \cite{mirza2014cGAN_original} and conditional variational autoencoders (cVAEs) \cite{sohn2015cVAE_original}. These models directly generate molecular structures to meet a set of given target conditions \cite{dong2020RCGAN, mendez2020cGAN, lim2018cVAE_seongokRyu, Kim2021GCT}. In contrast, there are also ways to obtain the desired materials from unconditional generative models, such as generative adversarial networks (GANs) \cite{goodfellow2014GAN_original} and variational autoencoders (VAEs) \cite{kingma2013VAE_original}. These approaches use additional methods to find appropriate latent code, which is required to generate the target-hitting substances. Navigating policies of latent space trained by reinforcement learning (RL) \cite{guimaraes201GAN+RL, sanchez2017GAN+RL} and optimization techniques \cite{blaschke2018VAE+BayesianOpt, griffiths2020VAE+BayesianOpt, long2021constrained} belong here.

Unfortunately, all of the previously mentioned models (NMT, GAN, and VAE-based) are difficult to use in materials extrapolation for discovering novel materials with properties that are out of training data distribution. To generate realistic molecules with these models, the models should be trained to generate molecular data that approximate the probability distribution of the real-world chemical system. However, since it is impossible to know the true probability of the real-world chemical system, the models are trained to generate data that approximate the empirical probability of the training data. Regrettably, the empirical data at our disposal may exhibit biases due to various factors, consequently leading to models trained on such biased data failing to generate some molecules that even exist in the real world (Fig.~\ref{fig:moses_distribution}a). Hence, the probability distribution-learning models are not suitable for generating molecules in regions with little or no known data (such as materials extrapolation). Furthermore, there are several ongoing discussions about whether probability distribution-learning models are suitable for extrapolation problems \cite{LOCO, ext_limit1, polykovskiy2020moses, mokaya2023testing}. In the same vein, we believe that employing an approach that either avoids using data or minimizes data usage, such as RL and genetic algorithm (GA), is appropriate for materials extrapolation. Since GA is a method for deriving a set of optimal solutions rather than a problem-solving policy, it cannot guarantee the diversity of the derived solution set. On the contrary, RL involves learning action policies to obtain solutions based on the given current state. This advantage enables RL to infer a wider variety of solutions. Therefore, we intend to utilize RL in our approach. Despite several recent studies utilizing RL in molecular design\cite{gottipati2020learning, reinvent, yang2021hit}, the majority of them do not prioritize presenting RL as a means to address the limitations of probability distribution-learning models for discovering substances beyond their trained domains.

Combinatorial chemistry \cite{combinatorial_chemistry} was invented in the 1980s and can generate molecules with properties out of known data. These types of methods use a set of molecular fragments and rules for fragment combination. Breaking of retrosynthetically interesting chemical substructures (BRICS) \cite{degen2008BRICS} is an example of combinatorial chemistry. This technique involves combining randomly selected BRICS fragments with their template-based fragment combination rules, which is similar to assembling Lego blocks. Therefore, combinatorial chemistry can create all chemically possible molecular structures that can be obtained from the combination of molecular fragments. According to our estimation, approximately $4\times 10^{16}$ types of small molecules ($\leq 500$ $Da$) can be combined with 2,207 BRICS fragments. Considering that the total number of small molecules was roughly estimated to be $10^{60}$ in ref. \cite{bohacek199610^60}, this means that it can cover a fairly wide area. However, there is the limitation that the combinatorial chemistry-based molecular generator does not know which molecular fragments to be selected and combined to complete the desired molecule. In other words, it has no policy to guide the selection of molecular fragments to obtain the target molecule. Hence, it proceeds with countless attempts to combine randomly selected fragments and selects the best compound from the generated molecular candidates, which can result in a combinatorial explosion \cite{klaus1986web}. If we assume that it takes 1 $ms$ to assemble one molecule, it would take $1.27\times10^{6}$ $years$ to enumerate all possible small organic molecules using 2,207 molecular fragments; $4\times10^{16}$ molecules $\times 1$ $ms$ $= 1.27\times10^{6}$ $years$.

Herein, we introduce RL to provide combinatorial chemistry with a molecular fragment selection policy that guides the generating molecule toward the target. With a randomly selected initial fragment, the RL-guided policy iteratively selects the subsequent fragment to be combined. In the training phase, the policy is learned by giving a higher reward if the properties of the generated molecule are closer to the target. Therefore, the learned policy enables an efficient search of chemical space and helps to escape from the combinatorial explosion problem by providing direction to the target. Moreover, the proposed model\textemdash \textit{RL-guided combinatorial chemistry (RL-CC)}\textemdash has the potential to enable materials extrapolation, which is impossible for probability distribution-learning models. To demonstrate the potential empirically, we apply RL-CC and two probability distribution-learning models to a toy problem of molecules discovery that hits multiple extreme target properties simultaneously. The results indicate that our model can discover extreme target molecules that probability distribution-learning models cannot reveal. Furthermore, we theoretically demonstrate why the probability distribution-learning models are not suitable for problems involving materials extrapolation. To illustrate the performance in actual problems, we conduct two practical experiments. The first is to discover protein docking molecules to a 5-hydroxytryptamine receptor 1B (5-HT\textsubscript{1B} receptor) with high binding affinity. The second is the discovery of human immunodeficiency virus (HIV) inhibitors with high potency. These two experiments demonstrate that the proposed approach can discover compounds with extreme properties, which shows the potential to be extended as materials extrapolation when it utilizes a set of domain-specific molecular fragments and their combination rules.

\section{Results and Discussion}\label{results}

\subsection{Theoretical review of probability distribution-learning models}\label{results_limit_of_dist}

Inverse molecular design models based on NMT, VAE, and GAN learn the empirical probability distribution of training data $P_{data}$. Let $X,Y=({X}_1, {Y}_1),...,({X}_N, {Y}_N)$ denote $N$-sampled training data. Here, ${X}_i=(x_{i,1},...,x_{i,T'})$ denotes sequence data of the $i$-th molecular structure, and ${Y}_i=(y_{i,1},...,y_{i,T})$ denotes a set of properties of the $i$-th molecule. The NMT-based models are trained to translate the input ${Y}_i=(y_{i,1},...,y_{i,T})$ into a paired output sequence ${X}_i=(x_{i,1},...,x_{i,T'})$. Here, $x_{i,t}$ is a one-hot encoded vector of the $t$-th token constituting a molecule ${X}_i$. The $\theta$-parameterized translator $G_\theta^{NLP}$ should be trained to select a token $x_{i,t}$ iteratively over $t=1,...,T'$, by maximizing the likelihood $\prod_{i=1}^{N}\prod_{t=1}^{T'} G_{\theta}^{NLP}\left(x_{i,t} \mid Y_i,x_{i,1:t-1}\right)$ empirically. The actual training process is conducted by minimizing its negative log-likelihood $-\sum_{i=1}^{N}\sum_{t=1}^{T'} \log G_{\theta}^{NLP}\left(x_{i,t} \mid Y_i,x_{i,1:t-1}\right)$, which is equivalent to minimizing cross-entropy $H(\cdot\,,\cdot)$ of hypothesis $\hat{X}_\theta$ from training data $X$:

\begin{equation} \label{eqn:loss_CE}
\begin{split}
H(X, \hat{X}_\theta) & = -\sum_{i=1}^{N} \sum_{t=1}^{T'} P(X) \log P(\hat{X}_\theta) \\
 & = H\left(X\right) + D_{KL}\left(P(X) \parallel P(\hat{X}_\theta)\right)
\end{split}
\end{equation}

\noindent where $H(X)=-\sum_{i=1}^{N} \sum_{t=1}^{T'} P(X)\log P(X)$ denotes the entropy \cite{shannon2001entropy} of training data X, and $D_{KL}(P(X) \parallel P(\hat{X}_\theta))$ is the Kullback–Leibler (KL) divergence \cite{kullback1951KLD} of hypothesis probability $P(\hat{X}_\theta)$ from $P_{data}$. Since $H(X)$ is not a function of trainable parameter $\theta$, minimizing the cross-entropy $H(X, \hat{X}_\theta)$ is equivalent to minimizing the KL divergence term in Equation (\ref{eqn:loss_CE}). Thus, the optimal $G_{\theta^*}^{NLP}$ is obtained by approximating $P(\hat{X}_\theta)$\textemdash which is the probability distribution of data generated by $G_{\theta}^{NLP}$\textemdash to $P_{data}$. It means that $G_{\theta}^{NLP}$ learns the empirical probability distribution of training data $P_{data}$, not the true probability distribution of the system $P$.

Second, VAE-based models are types of generative self-learning models that learn the empirical probability distribution of training data $P_{data}$. The models are trained to encode training data $X=X_1,...,X_N$ in the latent space with encoder $Q_{\phi}^{VAE}$ and reconstruct it with decoder $G_{\theta}^{VAE}$. The difference from autoencoders is that the latent variables $z$ are constrained to follow a prior distribution such as a normal distribution. The VAEs are trained using the following objective function \cite{kingma2013VAE_original}:

\begin{equation}
\label{eqn:negative-elbo}
\argmin_{\phi,\theta} \sum_{i=1}^{N}-\EX_{Q_{\phi}^{VAE}(z \mid X_i)}\left[\log G_{\theta}^{VAE}(X_i \mid z)\right] + D_{KL}\left(Q_{\phi}^{VAE}(z \mid X_i) \parallel P(z)\right)\\
\end{equation}

When looking at the VAEs from the perspective of a generative model, the KL divergence term simply acts like a regularizer in the training process. The expectation term measures the  reconstruction error of $G_{\theta}^{VAE}$ and it is approximated by using the following $L$-number of Monte-Carlo sampling:

\begin{equation}
\label{eqn:reformulated_negative-elbo}
-\EX_{Q_{\phi}^{VAE}(z \mid X_i)} \left[\log G_{\theta}^{VAE}(X_i \mid z)\right] \approx -\frac{1}{L}\sum_{z_j=1}^L \log\left(G_{\theta}^{VAE}(X_i \mid z_j)\right)\\
\end{equation}

Here, the approximated reconstruction error term is a form of negative log-likelihood over training data $X$. Hence, it is equivalent to minimizing the cross-entropy $H(\cdot\,,\cdot)$ of hypothesis $\hat{X}_{\theta,z}$ from training data $X$:

\begin{equation} \label{eqn:loss_CE_VAE}
\begin{split}
H(X, \hat{X}_{\theta,z}) & = -\sum_{i=1}^{N} \sum_{t=1}^{T'} P(X) \log P(\hat{X}_{\theta,z}) \\
 & = H\left(X\right) + D_{KL}\left(P(X) \parallel P(\hat{X}_{\theta,z})\right)
\end{split}
\end{equation}

As mentioned above, this is also equivalent to minimizing the KL divergence term in Equation (\ref{eqn:loss_CE_VAE}), since $H(X)$ is not a function of trainable parameter $\theta$. Hence, the optimal $G_{\theta^*}^{VAE}$ is obtained by approximating the probability distribution of hypothesis $P(\hat{X}_{\theta,z})$ to the $P_{data}$. This means that $G_{\theta}^{VAE}(X \mid z)$ with $z\sim Q_{\phi}^{VAE}\left(z \mid X\right)$ approximates $P_{data}$.

\begin{figure*}[h!]
    \centering
    \includegraphics[width=0.95\textwidth]{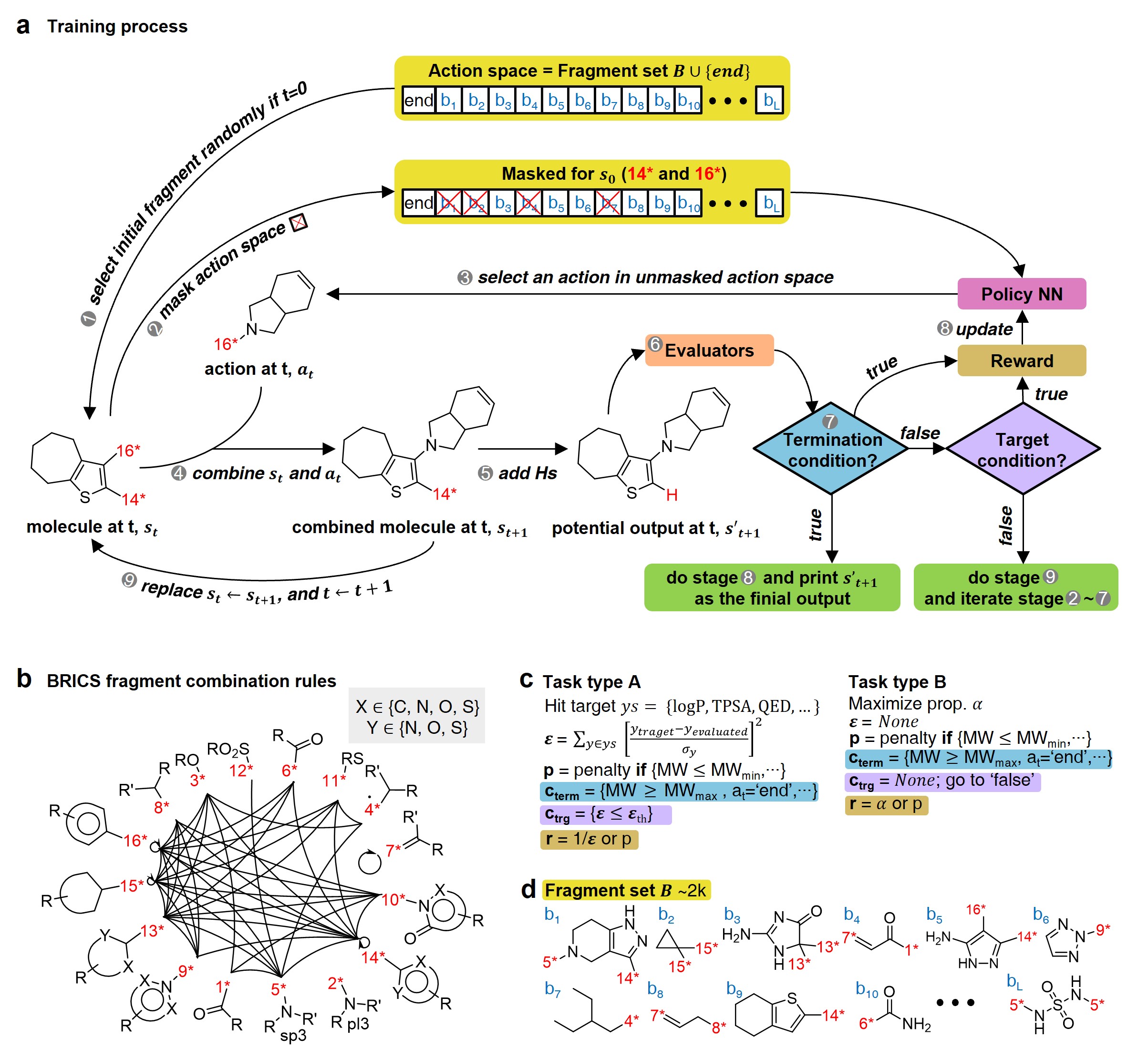}
    \caption{\textbf{Overview of RL-guided combinatorial chemistry with BRICS.} (a) Training process. (b) Modified BRICS \cite{degen2008BRICS} fragment combination rules. Here, the RDKit \cite{rdkit} version 2020.09.1.0 of the modified BRICS rules is adopted. This figure is modified from Degen et al., 2008, \emph{ChemMedChem}, 10(3): 1503-1507 \cite{degen2008BRICS}, with permission of Wiley-VCH GmbH. (c) Type of tasks. Task type A is to discover molecules that hit the specific values of given target properties and Task type B is to discover molecules that maximize the given target properties. (d) Fragment set $B$. Here, $B\cup\{end\}$ is defined as action space.}
    \label{fig:model_scheme}
\end{figure*}

Third, GAN is also a model to obtain a generator $G_{\theta}^{GAN}$ that approximates $P_{data}$. Here, $G_{\theta}^{GAN}$ learns the $P_{data}$ in the learning process to generate data that sufficiently resembles training data $X$ to deceive the discriminator $D_{\phi}^{GAN}$. Note that it has been proved that the global minimum of the virtual training criterion of the generator is achieved if (and only if) $P_{data} = G_{\theta}^{GAN}(z)$ \cite{goodfellow2014GAN_original}. This means that the optimal $G_{\theta^*}^{GAN}$ is obtained by approximating the hypothesis probability $P(\hat{X}_{z})$ to the probability of training data $P_{data}$.

Therefore, it can be concluded that models based on NMT, VAE, and GAN used for inverse molecular design are models to derive an approximator of $P_{data}$. Unfortunately, since $P_{data}$ derived from the observed empirical data is not equal to the true probability $P$ of chemical system, it cannot guarantee that the probability-distribution learning models will work well for problems involving materials extrapolation.

\subsection{RL-guided combinatorial chemistry with BRICS}\label{results_modeling}

\begin{figure*}[h!]
    \centering
    \includegraphics[width=0.95\textwidth]{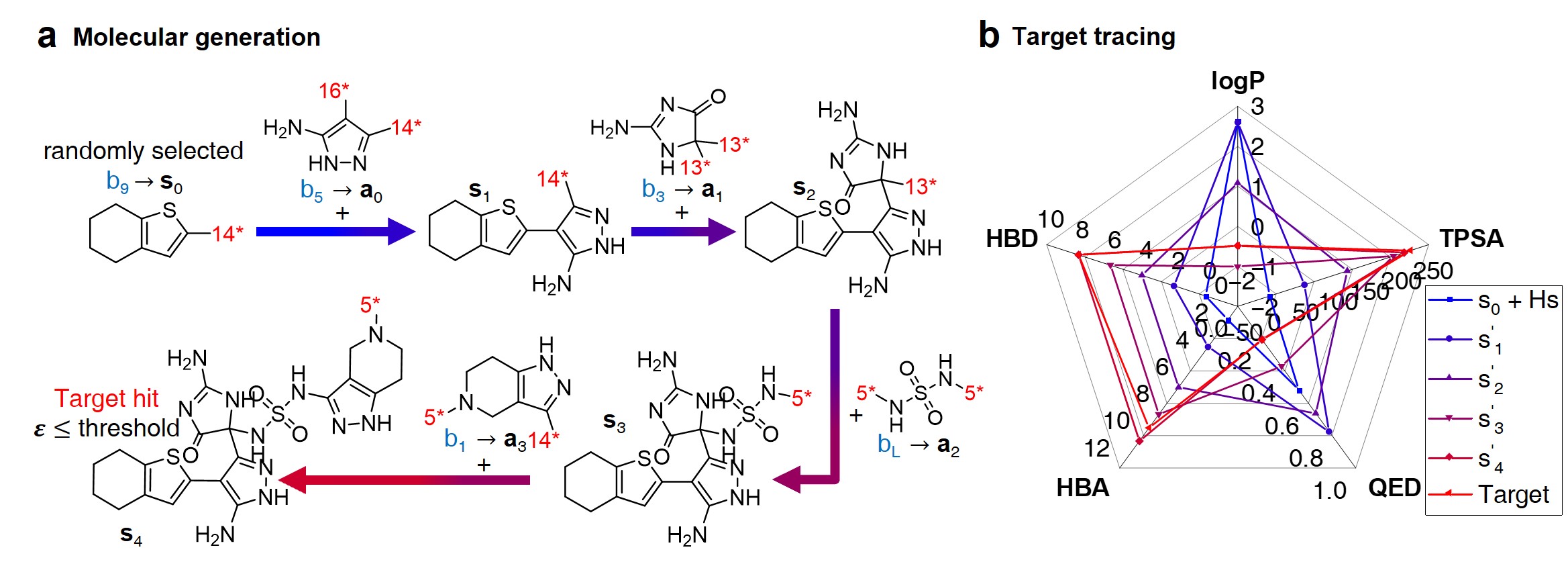}
    \caption{\textbf{Inference process for molecular generation.} (a) An example of a molecular generation process. (b) Property changes for generated molecules.
    }\label{fig:inference}
\end{figure*}

The RL-CC illustrated in Fig. \ref{fig:model_scheme} is applicable to various tasks of materials discovery, by designing a target molecule with the selected molecular fragments. A trained RL-guided policy iteratively selects the subsequent fragment to be combined. Here, the RL-guided policy serves as a guide to generate a target-hitting molecule. This approach has three main phases: configuration settings (Fig. \ref{fig:model_scheme}b--d), training (Fig. \ref{fig:model_scheme}a), and inference (Fig. \ref{fig:inference}).

In the configuration settings phase, all settings necessary for reinforcement learning are customized. Accordingly, the task for materials discovery must be specified. There are two types of tasks for this (Fig. \ref{fig:model_scheme}c): the discovery of molecules to hit specific values of the multiple target properties (Task type A), and the discovery of molecules to maximize a specific property (Task type B). Depending on the type of given task, the user designs the reward function $r$, target error function $\varepsilon$, termination conditions $c_{term}$, and target conditions $c_{trg}$. The reward function $r$ is designed to give a higher reward the better a given task is performed. For Task type A, the target error function $\varepsilon$ and reward function $r$ are designed as sum errors for the multiple target properties and the reciprocal of the target error function $\varepsilon$, respectively. In the case of Task type B, the property itself is used as the reward function $r$; hence, maximizing $r$ is equivalent to maximizing the property. For the minimization case of Task type B, the negative property is used as the reward function $r$. We can also consider the constraints $p$, which are reflected in the reward function $r$ by giving penalties if one of the constraints $p$ is not satisfied. Here, the minimum molecular weight ($MW_{min}$) and the minimum number of fragments ($n_{min}$) that make up a molecule can be used as constraints $p$. These enable the model to generate various molecules by preventing premature termination, which would cause the generation of molecules that were too small and uniform.

The termination conditions ($c_{term}$) and target conditions ($c_{trg}$) pertain to deciding when to terminate the process of selecting and combining additional molecular fragments, which determines the characters of the final output molecule. Hence, the termination conditions $c_{term}$ and target conditions $c_{trg}$ are designed considering the given task. The molecular generation process is terminated early if one of the termination conditions $c_{term}$ is satisfied. Accordingly, maximum molecular weight ($MW_{max}$) and maximum number of fragments ($n_{max}$) are used to design the termination conditions  $c_{term}$. It should be noted that the process is also terminated if there are no more sites for binding fragments to the combined molecule at step $t$ $(s_{t+1})$ or if the policy selects $end$-action at step $t$. These are included in the termination conditions $c_{term}$. The target conditions $c_{trg}$ are the target bounds to hit, which are applied only to Task type A.

To evaluate the previously mentioned functions and conditions, the evaluators (stage 6 in Fig. \ref{fig:model_scheme}a) for the interesting properties are utilized selectively to calculate the properties of potential output molecule at step $t$ $(s^{'}_{t+1})$. In order to calculate the properties of the molecule, the molecule must not have any unfilled binding site. Hence, $s^{'}_{t+1}$ is derived by attaching hydrogen to the unfilled binding sites of the combined molecule at step $t$ $(s_{t+1})$ if $s_{t+1}$ has any unfilled binding site (stage 5 in Fig. \ref{fig:model_scheme}a). RDKit \cite{rdkit} was selectively used to evaluate the calculated octanol-water partition coefficient (logP) \cite{logP}, topological polar surface area (TPSA) \cite{TPSA}, quantitative estimates of drug-likeness (QED) \cite{QED}, number of hydrogen bond acceptors (HBA), number of hydrogen bond donors (HBD), and molecular weight (MW). A quantitative structure-activity relationship (QSAR) model \cite{kotsias2020cRNN} and QVina2 \cite{alhossary2015fast}, which is a tool for discovering the minimum-energy docking conformation of a tested molecule and calculating its docking score, are also selectively used to evaluate drug activity for dopamine receptor D2 (DRD2) and the binding affinity to the 5-HT\textsubscript{1B} receptor, respectively.

For combinatorial chemistry, fragment set $B$ and its combination rules should be set. Accordingly, a modified version \cite{rdkit} of the BRICS \cite{degen2008BRICS} system was adopted (Fig. \ref{fig:model_scheme}b,d). Since the best performance was achieved for approximately 2k action spaces in the preliminary experiments (ESI\dag 
~Note 2), approximately 2k fragments were sampled from BRICS 40k for fragment set $B$.  The BRICS fragment combination rules are rules to bind 16 molecular templates, where each template has a unique binding site (red digit in Fig. \ref{fig:model_scheme}b).

In the training phase, our model was trained using the proximal policy optimization (PPO) algorithm \cite{schulman2017PPO}, which is known to perform well in RL problems with discrete actions. This is because it has the advantages of stable training, sample efficiency, scalability, and flexibility. In the preliminary experiments, PPO performed optimally for our problem among several state-of-the-art RL algorithms (ESI\dag 
~ Note 3). An episode iteratively proceeds the process of selecting and combining molecular fragments from steps 0 to $T$ (Fig. \ref{fig:model_scheme}a). If one of the termination conditions $c_{term}$ or target conditions $c_{trg}$ are satisfied, the episode is prematurely terminated at the step. At step 0, an episode is started with a randomly selected fragment $s_0$ (stage 1 in Fig. \ref{fig:model_scheme}a), and the randomness of the initial fragment $s_0$ creates more diverse output molecules $s^{'}_{T+1}$. Subsequently, action masking (stage 2 in Fig. \ref{fig:model_scheme}a) is performed, which masks the actions that are not applicable to $s_0$. Thereby, the fragment selection policy is enforced not to select the masked actions. This action masking helps to generate molecules that do not violate the chemical valence rule and enables efficient learning by reducing the action space. In this way, the fragment selection policy selects an action at step 0 ($a_0$) from the unmasked actions.

If the selected action at step $t$ ($a_t$) is the \textit{end}-action, the process is terminated. However, if $a_t$ is a fragment, $a_t$ is combined with $s_t$ to make a combined molecule at step $t$ ($s_{t+1}$). To evaluate the properties of a molecule, the molecule should not have any unfilled binding sites. Hence, the potential output molecule at step $t$ ($s^{'}_{t+1}$) is derived by attaching hydrogen to the remaining binding site of $s_{t+1}$. Then, the interesting properties of $s^{'}_{t+1}$ are evaluated to obtain the target error $\varepsilon_{t+1}$ and reward $r_{t+1}$ at step $t$. Now, to check whether one of the termination conditions $c_{term}$ or target conditions $c_{trg}$ is satisfied, $a_t$, $\varepsilon_{t+1}$, and $r_{t+1}$ are considered. If one of the termination conditions $c_{term}$ or target conditions $c_{trg}$ is satisfied, the reward $r_{t+1}$ is used to update the policy. If any of the termination conditions $c_{term}$ and target conditions $c_{trg}$ is not satisfied, the environment outputs the combined molecule $s_{t+1}$. Then, the model iteratively proceeds to the next step of the process until either one of the termination conditions $c_{term}$ or target conditions $c_{trg}$ is satisfied. This process is repeated for a preset number of iterations to train the model.

After the training is completed, the trained policy is used to generate a target molecule in the inference phase. Fig. \ref{fig:inference}a displays an example of molecular generation, in which new molecular fragments are selected and combined to complete a target molecule from steps 0 to 3. In the process, the properties of the generated potential output molecules ($s^{'}_1$ to $s^{'}_4$), which are derived from hydrogen attachment of the combined molecules ($s_1$ to $s_4$), change from the properties of hydrogen attached initial fragment ($s_0 + Hs$) to the target properties (logP: $-0.488$, TPSA: $220.83$, QED: $0.207$, HBA: $9$, HBD: $8$). In step 3, the properties of the potential output molecule $s^{'}_4$ (logP: $-0.488$, TPSA: $211.09$, QED: $0.205$, HBA: $10$, HBD: $8$) are close to the target properties (Fig. \ref{fig:inference}b). Since the target error $\varepsilon_4$ is lower than the maximum target error $\varepsilon_{max}$, the process is early terminated at step 3. Hence, the potential output molecule at step 3 ($s^{'}_{4}$) becomes the final output molecule.

\subsection{Materials extrapolation to hit multiple extreme target properties}\label{results_chembl}

In this section, we empirically demonstrate that RL-guided combinatorial chemistry enables the discovery of extrapolated compounds, which is not possible with probability distribution-learning models. To achieve this, we adopt two different types of probability distribution-learning models and compare their performance with our model in terms of materials extrapolation. One of the adopted models is a conditional recurrent neural network (cRNN) \cite{kotsias2020cRNN}. It serves as an NMT-based translator, translating input target properties into the corresponding molecular chemical language. As it operates as a translator, the input target properties and the corresponding molecular chemical language are one-to-one matched. The other model is the generative chemical Transformer (GCT)\cite{Kim2021GCT}, a cVAE-based generative model. It utilizes the Transformer’s architecture as the backbone and incorporates a conditional latent space between its encoder and decoder. GCT generates chemical language corresponding to input target properties and sampled noise. As it functions as a generator using randomly sampled noise, it generates various molecular chemical languages with a single set of target properties.

For the demonstration, we conducted experiments on generating molecules to hit multiple target properties. The experimental setup was borrowed from ref. \cite{kotsias2020cRNN}. In the experiments, the following seven drug-related target properties were given to the models to generate target-hitting molecules: logP, TPSA, QED, HBA, HBD, MW, and DRD2. This experiment falls under Task type A as specified in Fig.\ref{fig:model_scheme}c. Since $\epsilon$ has multiple terms to minimize and multiple constraints, this toy problem covers complex optimization problems. By even changing the signs of some terms of $\epsilon$, it is possible to cover complex optimization problems that involve a mix of minimization and maximization. Detailed information about the properties is summarized in Methods. Here, RDKit and a QSAR model for DRD2 \cite{kotsias2020cRNN} were adopted as evaluators.

\begin{table}[t]
\centering
\caption{The target-hitting errors for materials interpolation}
\label{tab:chembl_interpolation_error}
\begin{threeparttable}
\begin{tabular}{lccc} 
\toprule
 & \multicolumn{2}{c}{$RMSE_i$} & \multicolumn{1}{c}{$\overline{RMSE_i}^{a}$} \\
\cmidrule(lr){2-3} \cmidrule(lr){4-4}
& cRNN \cite{kotsias2020cRNN} & GCT \cite{Kim2021GCT} & Average \\
\midrule
logP & 0.379 & 0.368 & 0.373 \\
TPSA & 5.476 & 5.109 & 5.292 \\
QED & 0.081 & 0.075 & 0.078 \\
HBA & 0.932 & 1.204 & 1.068 \\
HBD & 0.223 & 0.247 & 0.235 \\
MW & 5.954 & 8.272 & 7.113 \\
DRD2 & 0.113 & 0.098 & 0.105 \\
\bottomrule
\end{tabular}
\end{threeparttable}
    \begin{tablenotes}
    \sloppy
        \item[a] $^a\overline{RMSE_i}$ refers average $RMSE$ of target property $i\in \{logP, TPSA, QED, HBA, HBD, MW, DRD2\}$ for cRNN \cite{kotsias2020cRNN} and GCT \cite{Kim2021GCT}.
    \fussy
    \end{tablenotes}
\end{table}

First, the original performances of cRNN \cite{kotsias2020cRNN} and GCT \cite{Kim2021GCT} in interpolation points were evaluated (Table \ref{tab:chembl_interpolation}). The two models were trained and tested with datasets \cite{kotsias2020cRNN} curated from the ChEMBL database \cite{mendez2019chembl}. Here, how well target-hitting molecules were generated for the given target properties\textemdash which were gathered from 149,679 molecules in the curated ChEMBL test data \cite{kotsias2020cRNN}\textemdash was evaluated. In contrast to the two methodologies, the proposed RL-CC approach in this paper requires retraining for each set of target properties. This makes it impractical for a realistic performance comparison in terms of interpolation, as training needs to be conducted individually for 149,679 properties. Additionally, RL-CC focuses on generating molecules with extreme properties, and thus, model performance in interpolation was not assessed for RL-CC.

\begin{figure*}[!t]
    \centering
    \includegraphics[width=0.95\textwidth]{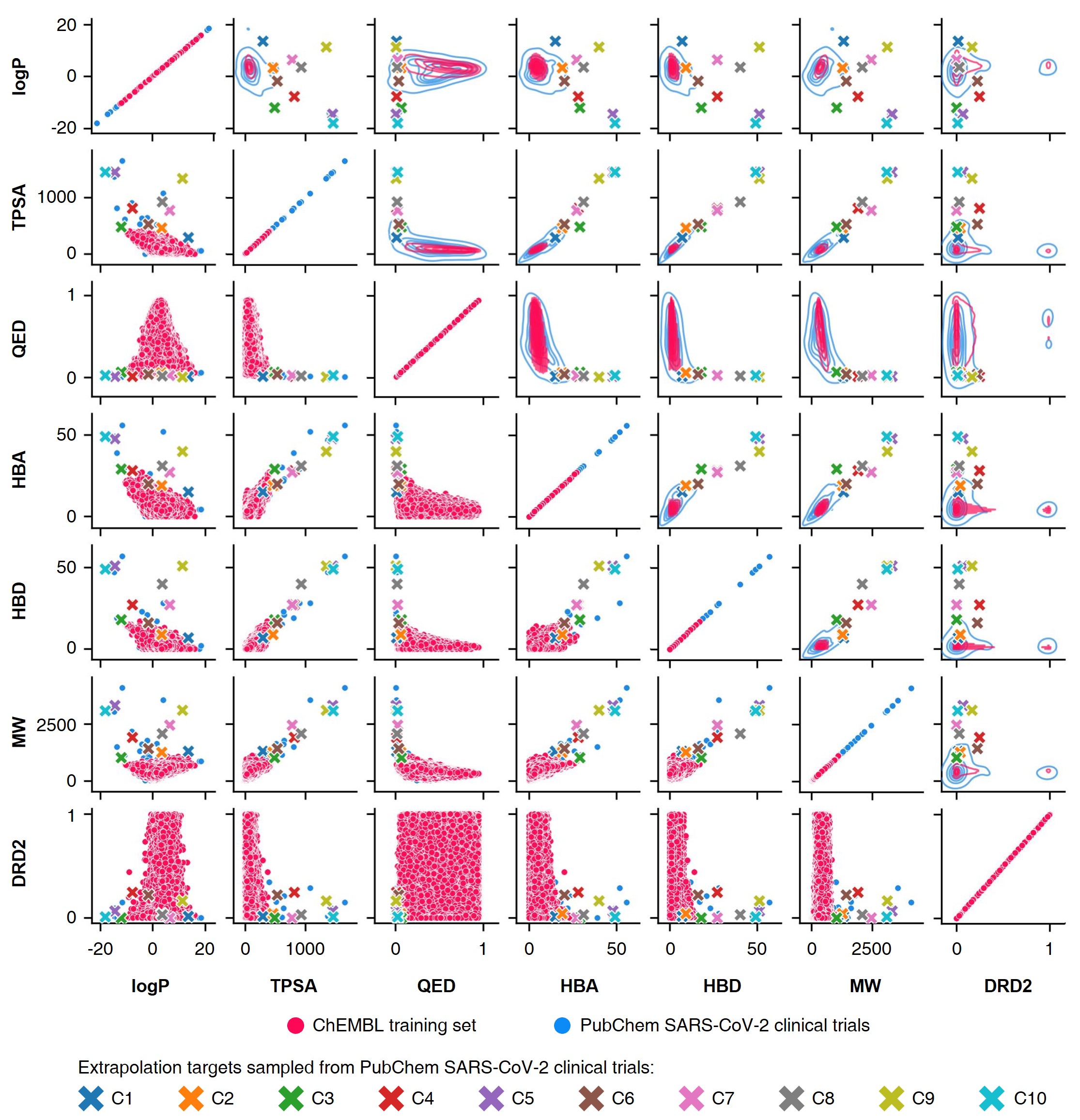}
    \caption{\textbf{Targets for materials extrapolation.} The PubChem SARS-CoV-2 clinical trials dataset \cite{PubChemCovid19} is more widely distributed than the ChEBML training dataset \cite{mendez2019chembl}. The properties of five molecules in PubChem SARS-CoV clinical trials that deviated from the logP-TPSA distribution of the ChEMBL training dataset were set as extrapolation targets C1 to C5, and the properties of five molecules that deviated from the TPSA-QED distribution were set as extrapolation targets C6 to C10.
    }\label{fig:chembl_results}
\end{figure*}

To conduct the experiments on materials extrapolation, we adopted another molecular dataset with properties that were more widely distributed than the trained data \cite{kotsias2020cRNN}: PubChem SARS-CoV-2 clinical trials \cite{PubChemCovid19} (Fig. \ref{fig:chembl_results}). Among the molecules in this dataset, 10 molecules were sampled whose properties were outside the trained data, which were then set as the extrapolated targets (C1 to C10). Since these 10 molecules were real molecules that exist in the real world, their properties would be physically feasible targets to generate. For each extrapolated target, 10,000 molecular generations were tried with cRNN \cite{kotsias2020cRNN} and GCT \cite{Kim2021GCT}.

To evaluate the performance of generating molecules that hit multiple target properties, the criteria for determining whether each target property was hit should be defined. Accordingly, in the experiment of materials interpolation with cRNN \cite{kotsias2020cRNN} and GCT \cite{Kim2021GCT}, the root mean squared error of each target property $i$ ($RMSE_{i}$) was analyzed (Table \ref{tab:chembl_interpolation_error}). Since all $RMSE_{i}$ for cRNN and GCT were not significantly different from each other, we determined that the magnitude of the average $RMSE_{i}$ ($\overline{RMSE_{i}}$) represents the difficulty of generating molecules that hit the target property $i$. Therefore, by setting $target$ $i\pm \overline{RMSE_{i}}$ as the target bound of $i$, a wide bound was assigned to targets that were difficult to hit and a narrow bound was assigned to any targets that were easy to hit. In this context, the term ‘target-hitting molecule’ refers to molecules with properties that fall within the range of $\pm\overline{RMSE_{i}}$ from the target values for each property.

\begingroup
\renewcommand{\arraystretch}{0.3} 

\begin{table*}[!t]
\centering
\caption{Performance benchmark for materials interpolation}\label{tab:chembl_interpolation}
\resizebox{0.85\linewidth}{!}{%
\begin{tabular}[t]{c|c|c|c|c|c|c|c|c|c|c}
\toprule
\rotatebox{90}{Model}
& \rotatebox{90}{\parbox{3cm}{\# of valid mols.\\ (\# of unique mols.)}}
& \rotatebox{90}{\parbox{4cm}{\# of all target-hitting mols.\\ w/o MW \& DRD2 \\(\# of unique mols.)}}
& \rotatebox{90}{\parbox{4cm}{\# of all target-hitting mols.\\ (\# of unique mols.)}}
& \multicolumn{7}{c}{\# of each target-hitting mols.}
\\
& & & & logP & TPSA & QED & HBA & HBD & MW & DRD2\\
\midrule

\rotatebox{90}{cRNN\cite{kotsias2020cRNN}}
& 8475 (8475) & 3774 (3774) & 2948 (2948) & 6263 & 6459 & 6653 & 7627 & 8077 & 6897 & 7796\\

\hline
\rotatebox{90}{GCT\cite{Kim2021GCT}}
& 8715 (8715) & 3480 (3436) & 2321 (2290) & 6289 & 6916 & 6693 & 7013 & 8200 & 5890 & 8114\\

\bottomrule
\end{tabular}
}

\end{table*}

\endgroup

\begingroup
\renewcommand{\arraystretch}{0.7} 
\begin{table*}[!h]
\centering
\caption{Performance benchmark for materials extrapolation}\label{tab:chembl_extrapolation}
\resizebox{0.9\linewidth}{!}{%
\begin{tabular}[t]{c|c|c|c|c|c|c|c|c|c|c|c}
\toprule
\rotatebox{90}{Model}
& \rotatebox{90}{Extrapolated targets}
& \rotatebox{90}{\parbox{4cm}{\# of valid mols.\\ (\# of unique mols.)}}
& \rotatebox{90}{\parbox{5cm}{\# of all target-hitting mols.\\ w/o MW \& DRD2 \\(\# of unique mols.)}}
& \rotatebox{90}{\parbox{5cm}{\# of all target-hitting mols.\\ (\# of unique mols.)}}
& \multicolumn{7}{c}{\# of each target-hitting mols.}
\\
& & & & & logP & TPSA & QED & HBA & HBD & MW & DRD2\\
\midrule
\multirow{10}{*}[0ex]{\rotatebox{90}{RL-CC (This work)}}
 &C10 & 10000 (9999)  & 23 (23)     & 0 (0)     & 820  & 1133 & 9978 & 5533 & 2618 & 0    & 35  \\
 &C9  & 10000 (10000) & 289 (289)   & 0 (0)     & 3919 & 1432 & 9995 & 7880 & 3223 & 1    & 15  \\
 &C8  & 10000 (9998)  & 181 (181)   & 0 (0)     & 2999 & 974  & 9992 & 6273 & 2379 & 466  & 133 \\
 &C7  & 10000 (9996)  & 321 (321)   & 0 (0)     & 3686 & 1420 & 9971 & 7832 & 3588 & 531  & 28  \\
 &C6  & 10000 (9981)  & 1709 (1708) & 311 (311) & 3769 & 3450 & 9890 & 8636 & 7301 & 1445 & 8729 \\
 &C5  & 10000 (10000) & 14 (14)     & 0 (0)     & 1607 & 1377 & 9981 & 7381 & 2363 & 0    & 26\\
 &C4  & 10000 (9969)  & 439 (439)   & 50 (50)   & 1972 & 1452 & 9809 & 7283 & 4384 & 572  & 9493\\
 &C3  & 10000 (9869)  & 913 (877)   & 233 (224) & 3143 & 2296 & 9940 & 8392 & 7143 & 1128 & 8310\\
 &C2  & 10000 (9987)  & 2424 (2424) & 366 (366) & 6738 & 3145 & 9928 & 8415 & 8635 & 1388 & 4970\\
 &C1 & 10000 (9994)  & 1317 (1317) & 355 (355) & 5202 & 2282 & 9918 & 9347 & 8622 & 1849 & 9406\\
\hline
\multirow{10}{*}[0ex]{\rotatebox{90}{cRNN\cite{kotsias2020cRNN}}}
 &C10 & 422 (2)  & 0 (0) & 0 (0) & 0 & 0 & 0 & 0 & 0 & 0 & 422\\
 &C9  & 0 (0)    & 0 (0) & 0 (0) & 0 & 0 & 0 & 0 & 0 & 0 & 0\\
 &C8  & 0 (0)    & 0 (0) & 0 (0) & 0 & 0 & 0 & 0 & 0 & 0 & 0\\
 &C7  & 0 (0)    & 0 (0) & 0 (0) & 0 & 0 & 0 & 0 & 0 & 0 & 0\\
 &C6  & 6 (6)    & 0 (0) & 0 (0) & 0 & 0 & 6 & 1 & 0 & 0 & 5\\
 &C5  & 0 (0)    & 0 (0) & 0 (0) & 0 & 0 & 0 & 0 & 0 & 0 & 0\\
 &C4  & 9836 (1) & 0 (0) & 0 (0) & 0 & 0 & 0 & 0 & 0 & 0 & 0\\
 &C3  & 3022 (5) & 0 (0) & 0 (0) & 0 & 0 & 4 & 2 & 1 & 0 & 3022\\
 &C2  & 0 (0)    & 0 (0) & 0 (0) & 0 & 0 & 0 & 0 & 0 & 0 & 0\\
 &C1  &1788 (10) & 0 (0) & 0 (0) & 0 & 0 & 4 & 1 & 1 & 0 & 1788\\
\hline
\multirow{10}{*}[0ex]{\rotatebox{90}{GCT\cite{Kim2021GCT}}}
 &C10 & 0 (0) & 0 (0) & 0 (0) & 0 & 0 & 0 & 0 & 0 & 0 & 0\\
 &C9  & 0 (0) & 0 (0) & 0 (0) & 0 & 0 & 0 & 0 & 0 & 0 & 0\\
 &C8  & 0 (0) & 0 (0) & 0 (0) & 0 & 0 & 0 & 0 & 0 & 0 & 0\\
 &C7  & 0 (0) & 0 (0) & 0 (0) & 0 & 0 & 0 & 0 & 0 & 0 & 0\\
 &C6  & 0 (0) & 0 (0) & 0 (0) & 0 & 0 & 0 & 0 & 0 & 0 & 0\\
 &C5  & 0 (0) & 0 (0) & 0 (0) & 0 & 0 & 0 & 0 & 0 & 0 & 0\\
 &C4  & 0 (0) & 0 (0) & 0 (0) & 0 & 0 & 0 & 0 & 0 & 0 & 0\\
 &C3  & 3 (3) & 0 (0) & 0 (0) & 0 & 0 & 3 & 1 & 0 & 0 & 3\\
 &C2  & 0 (0) & 0 (0) & 0 (0) & 0 & 0 & 0 & 0 & 0 & 0 & 0\\
 &C1  & 4 (4) & 0 (0) & 0 (0) & 0 & 0 & 1 & 0 & 0 & 0 & 4\\
\bottomrule
\end{tabular}
}

\end{table*}

\endgroup

Table \ref{tab:chembl_interpolation} shows the results for materials interpolation. In Table \ref{tab:chembl_interpolation}, the results of interpolation for a total of 149,679 target properties were rescaled to 10,000 for comparison with the extrapolation results. Since the number of attempted molecular generations for materials interpolation and extrapolation were different, the rescaled results based on 10,000 trials are summarized in Table \ref{tab:chembl_interpolation} for comparison. For materials interpolation, the cRNN \cite{kotsias2020cRNN} generated 2,948 molecules that hit all of the seven targets (logP, TPSA, QED, HBA, HBD, MW, and DRD2) simultaneously and 3,774 molecules hit the five targets (logP, TPSA, QED, HBA, and HBD) simultaneously. With the same criteria, the GCT \cite{Kim2021GCT} generated 2,321 molecules that hit all seven targets and 3,480 molecules hit the five targets simultaneously. These results confirmed that both models are able to generate target-hitting molecules in the trained region.

\begin{figure*}[!h]
    \centering
    \includegraphics[width=0.95\textwidth]{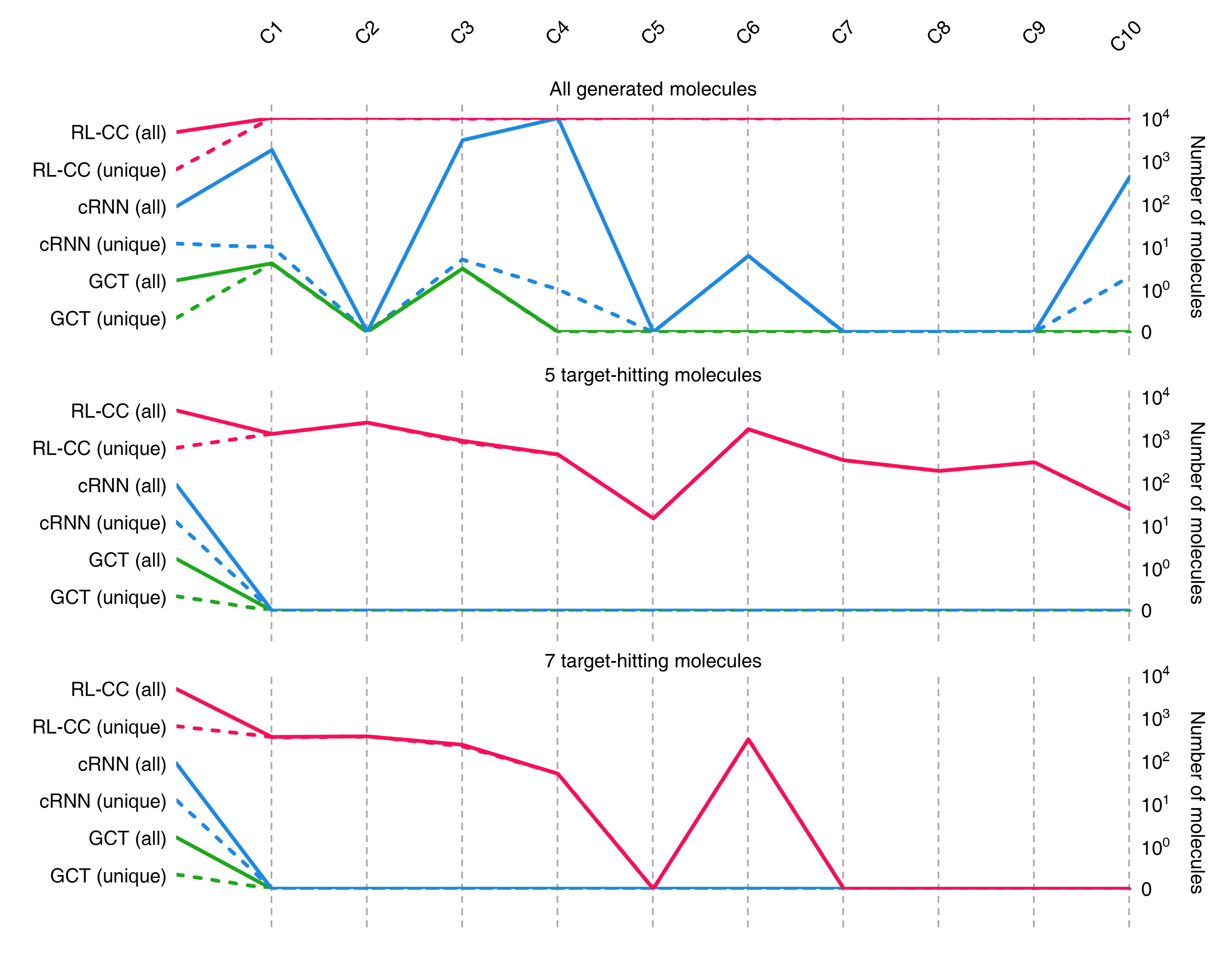}
    \caption{\textbf{Quality benchmarks of generated molecules in materials extrapolation.} Number of chemically valid molecules, 5 target-hitting molecules (logP, TPSA, QED, HBA, and HBD), and 7 target-hitting molecules(logP, TPSA, QED, HBA, HBD, MW, and DRD2)} out of10,000 molecules produced for target C1 to C10. A solid line indicates the total number of valid molecules meeting each condition, and a short-dashed line shows the number of unique valid molecules (excluding overlaps).\label{fig:sa_results}
\end{figure*}

Results for materials extrapolation are shown in Table \ref{tab:chembl_extrapolation} and Fig. \ref{fig:sa_results}. According to the results, both probability distribution-learning models achieved poor results in terms of molecular extrapolation. For each target from C1 to C10 outside the trained data, we conducted 10,000 trials to generate the molecules per target. As shown in the results for the targets C1 to C10 of Table \ref{tab:chembl_extrapolation} and Fig. \ref{fig:sa_results}, both probability distribution-learning models failed to generate molecules that hit all of the targets and did not generate molecules that hit the five targets. GCT \cite{Kim2021GCT} only succeeded in generating seven valid and unique molecules that satisfied the chemical valence rule out of 100,000 trials, with four and three molecules being generated for targets C1 and C3, respectively. cRNN \cite{kotsias2020cRNN} generated a total of 15,068 chemically valid molecules, although only 21 were unique. In particular, 15,054 out of the 15,068 valid molecules were nonsensical outcomes that were overlapped and too small, such as $CH_4$, $CH_4S$, $H_2O$, $H_2S$, $SO$, $H_2OS$, and $H_2S_2$. The MWs of these small molecules range from 16 and 66 $Da$. Considering that the target MWs of C1 to C10 ranged from 1,026 to 3,124 $Da$, it was difficult to conclude whether it operated correctly. Moreover, other generated molecules exhibited considerable deviations from the intended targets. Detailed information on the generated molecules is summarized in Table S1--S4\dag.

For the targets C1, C2, C3, C4, and C6, our RL-guided combinatorial chemistry generated a total of 1,315 target-hitting molecules that hit all seven targets simultaneously. Here, 355, 366, 233, 50, and 311 molecules that hit all targets were generated for targets C1, C2, C3, C4, and C6, respectively. For targets C5, C7, C8, C9, and C10, RL-guided combinatorial chemistry could not generate molecules that hit all seven target properties. However, it successfully generated a total of 828 molecules that hit five target properties that failed with the probability distribution-learning models. Here, 14, 321, 181, 289, and 23 molecules that hit five targets were generated for targets C5, C7, C8, C9, and C10, respectively.

Also, in fact, it is hard to assert that it completely failed to hit the seven target properties simultaneously for the targets C5, C7, C8, C9, and C10. For these targets, the generated molecules exhibited low target-hitting errors. This means that if the target bounds were more broadly set, there would be more molecules that were counted as molecules that hit all targets. It should be noted that the employed MW target bound $\pm$7.113 $Da$ and DRD2 target bound $\pm$0.105 were fairly narrow (See Table \ref{tab:chembl_interpolation_error}). The target bound of MW 7.113 $Da$ was so small that it was less than the weight of a single atom. Furthermore, the target bound of DRD2 0.105 was considerably smaller than the commonly known drug activity prediction accuracy of QSAR models. In ref. \cite{kotsias2020cRNN}, the QSAR model for DRD2 was used as a binary classifier to evaluate as either active (when the predicted value was $>$ 0.5) or inactive ($\leq$ 0.5). For this reason, we believe that the number of molecules hitting the targets MW and DRD2 was less counted than the number of molecules hitting the other targets (See MW-column and DRD2-column of RL-CC in Table \ref{tab:chembl_extrapolation}). Hence, we conducted a supplementary experiment on another dataset to generate five target-hitting molecules, excluding MW and DRD2 (ESI\dag 
~Note 1). As a result, we confirmed that our model successfully generated molecules with extreme properties outside the known domain, which is not possible with probability distribution-learning models.

Moreover, RL-CC uses a selective binding approach within combinable molecular fragments based on the BRICS fragment combination rules. As a result, all generated molecules adhered to the chemical valence constraints, ensuring a 100\% chemically valid. Furthermore, even when extreme target properties were specified, the diversity of chemically valid generated molecules remained remarkably high as seen in Fig. \ref{fig:sa_results}. 

When choosing the molecular candidates, it is also important to consider the synthesizability of the generated molecules. Thus, we have analyzed the synthetic accessibility (SA) score of the generated molecules using RDKit, which represents the ease of chemical compounds to be synthesized or produced. SA score is evaluated by considering how common the fragments composing the molecule are and the complexity of intricate ring structures such as fused rings. The SA scores for the generated molecules are presented in Figure S6\dag. The calculated results indicate that the generated molecules consistently exhibit low SA scores, excluding cases with extremely high molecular weights as seen in the experiment for generating molecules with extreme target properties. This suggests a reasonable level of synthesizability for the generated molecules.

\subsection{Application to the discovery of protein docking molecules}\label{subresults5}

The discovery of small molecules that dock to a target protein is a practical problem in drug discovery. Moreover, binding affinity to a target receptor is an important indicator for measuring drug-target interactions \cite{kitchen2004docking}. Since molecules with higher binding affinity to the target protein (compared to other proteins) can be considered as having high selectivity and docking ability, the discovery of molecules that maximize target binding affinity is a key objective in protein docking drug discovery \cite{binding_affinity}. Therefore, to generate molecules with a low docking score, which means to generate molecules that can bind strongly with 5-HT\textsubscript{1B} in this case, the trained policy attaches fragments that can maximize the reward.

\begin{figure*}[!h]
    \centering
    \includegraphics[width=0.95\textwidth]{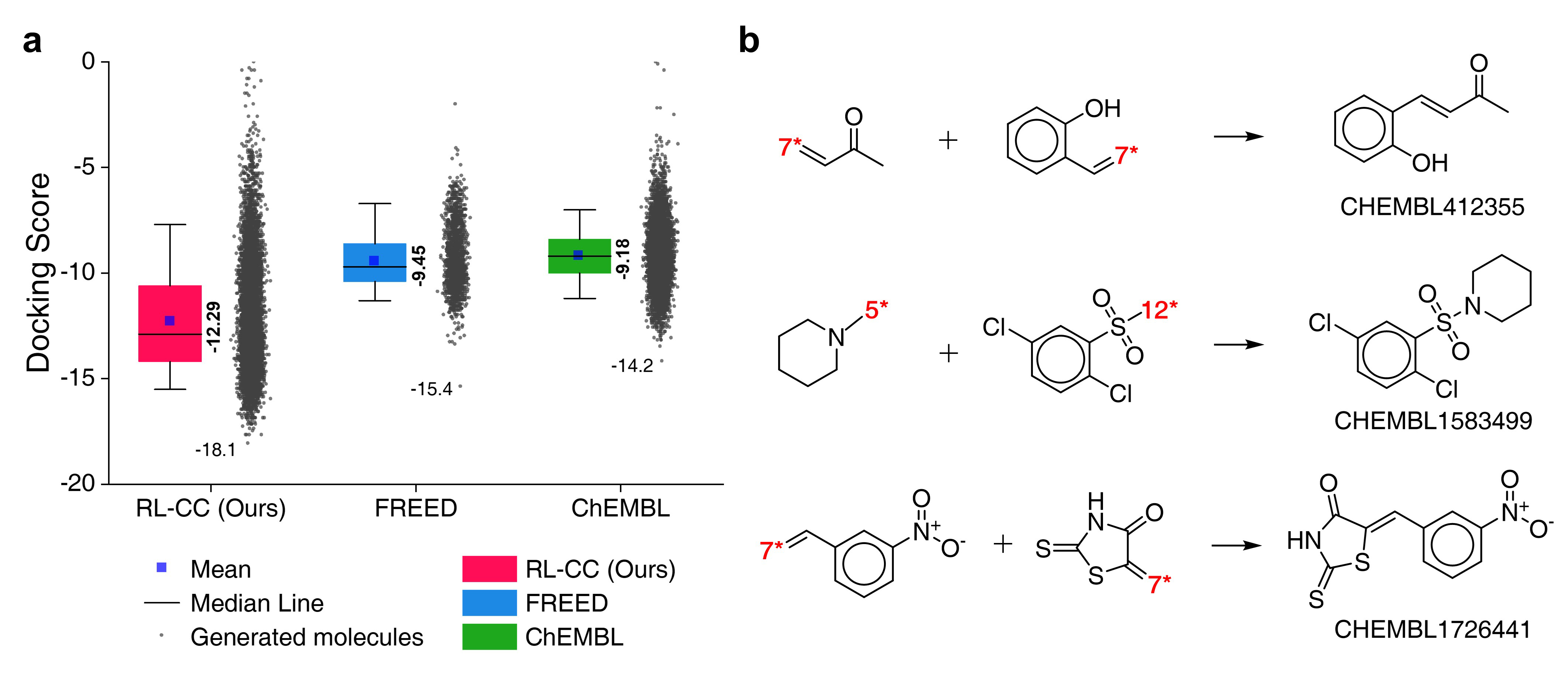}
    \caption{\textbf{Generated 5-HT\textsubscript{1B} receptor docking materials.} (a) Comparison of docking scores for the three molecular sets. Pink indicates the docking scores of 10,000 generated molecules from RL-guided combinatorial chemistry. Green indicates the docking scores of 10,000 drug-like molecules that were randomly sampled from ChEMBL \cite{mendez2019chembl} database. Blue indicates the docking scores of 1,871 molecules generated by FREED \cite{yang2021hit}. The 1,871 molecules have been reported as de novo cases with 4-step in the paper. It should be noted that the maximum number of fragments constituting a compound was the same as ours. The docking scores of the 1,871 molecules were re-evaluated using QVina2 \cite{alhossary2015fast} under the same calculation configuration as ours. The mean and minimum docking score for each outcome are represented numerically, while the box is set with percentiles 25 and 75, and the whiskers extend to the 5\textsuperscript{th} and 95\textsuperscript{th} percentiles. (b) Three molecular examples that were generated by RL-guided combinatorial chemistry, which exactly matched with active drug molecules reported in the PubChem Bioassay database \cite{wang2012pubchem}.}\label{fig:docking_results}
\end{figure*}

In this section, we demonstrate that RL-guided combinatorial chemistry can discover molecules that maximize the binding affinity towards the 5-hydroxytryptamine receptor 1B (5-HT\textsubscript{1B} receptor), which is related to mental diseases. A detailed description of the 5-HT\textsubscript{1B} receptor is summarized in Methods. We adopted QVina2 \cite{alhossary2015fast} (a data-free molecular docking simulator) to discover the minimum-energy docking conformation. This simulator evaluates the docking score quickly and reliably. Since the docking score is an indicator that is inversely proportional to the binding affinity, the reward function was set as the negative docking score.

\begin{figure*}[!h]
    \centering
    \includegraphics[width=0.95\textwidth]{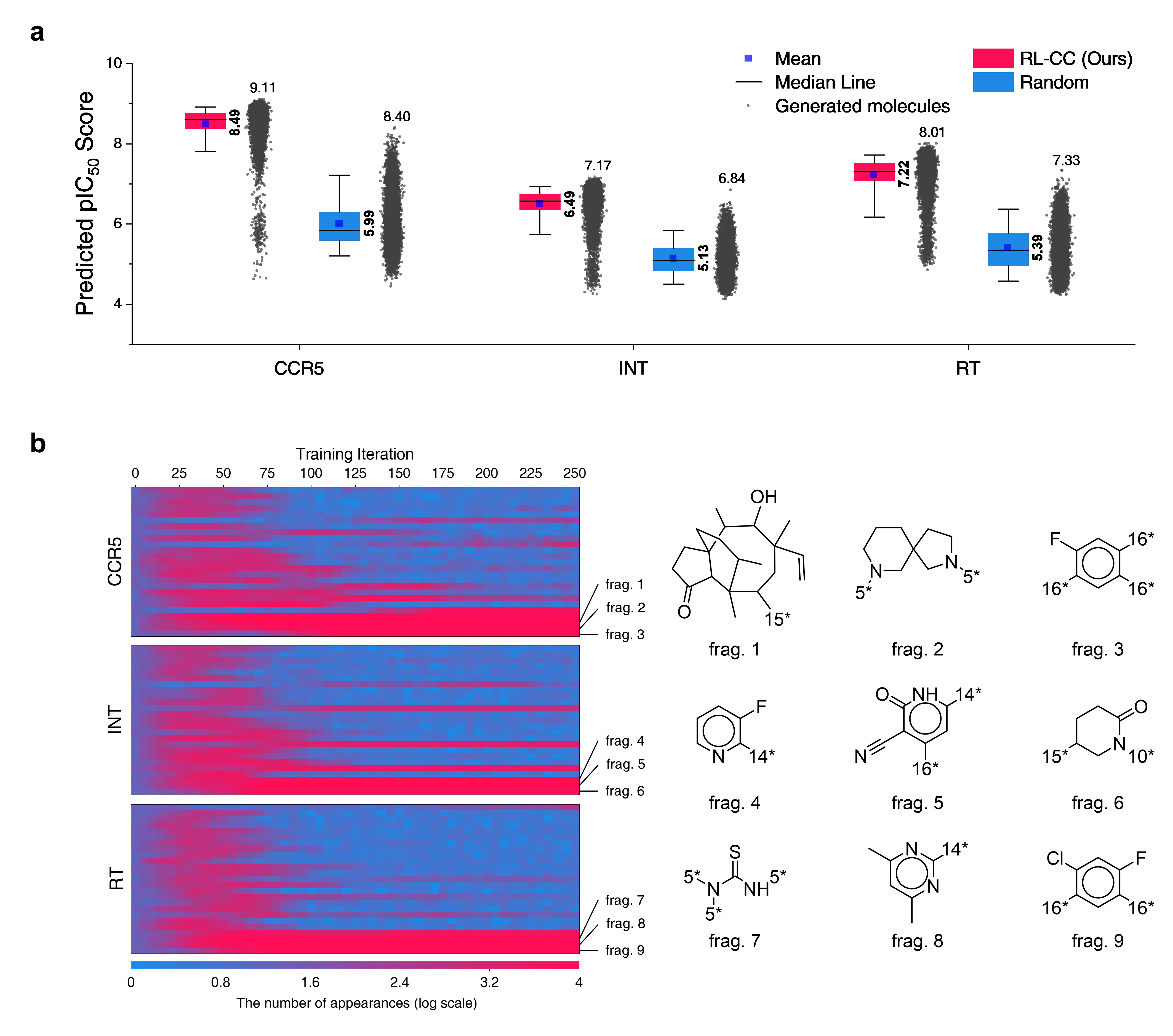}
    \caption{\textbf{Results for HIV inhibitors discovery with high pIC\textsubscript{50} for three HIV inhibition targets: CCR5, INT, and RT.} (a) Comparison of pIC\textsubscript{50} for two molecular sets. Pink indicates pIC\textsubscript{50} for a set of 10,000 molecules generated by RL-guided combinatorial chemistry. Blue indicates pIC\textsubscript{50} for a set of 10,000 drug-like molecules that were randomly sampled from the ChEMBL database \cite{mendez2019chembl}. The mean and maximum pIC\textsubscript{50} values for each outcome are represented numerically, while the box is set with percentiles 25 and 75, and the whiskers extend to the 5\textsuperscript{th} and 95\textsuperscript{th} percentiles. (b) Policy changes in BRICS molecular fragment selection according to the training steps. The left-hand side of the figure shows the number of appearances for 25 molecular fragments with the biggest change. The vertical and horizontal axes of the left-hand side blue-red plot represent the type of fragment and the training iteration, respectively. Blue-red indicates the number of generated molecules that have the fragment among the 10,000 generated molecules.}
    \label{fig:HIV_results}
\end{figure*}

To evaluate the performance of RL-guided combinatorial chemistry, the docking scores of 10,000 generated molecules from our model were compared with the docking scores of two other molecular sets. One was a set of 1,871 molecules that were generated to maximize the negative docking score towards the 5-HT\textsubscript{1B} receptor using fragment-based generative RL with explorative experience replay for drug design (FREED) \cite{yang2021hit}, which is reported in the paper. The other set was 10,000 molecules that were randomly sampled from ChEMBL \cite{mendez2019chembl} drug-like molecules. The docking scores of the three sets are summarized in Fig. \ref{fig:docking_results}a, which were calculated with QVina2. The calculation configuration is described in ESI\dag ~Note 4. The best molecule with the lowest docking score was discovered from our model and the median docking score for the sets was also the lowest. The top 10 generated molecular structures with the highest pIC\textsubscript{50} for CCR5, INT, and RT are illustrated in Fig. S8--S10\dag.

To check if potential drug molecules were found among the 10,000 generated molecules, we investigated whether any generated molecules were an exact match with the drug-like molecules in the ChEMBL \cite{mendez2019chembl} database. There were 23 molecules whose molecular structures exactly matched with real drug-like molecules in the ChEMBL database, of which 13 out of the 23 molecules had labels on drug activity (active or inactive). It should be noted that ref. \cite{bohacek199610^60} roughly estimated the number of small organic molecules as $10^{60}$, of which only $2.2\times 10^6$ were included in the ChEMBL database \cite{mendez2019chembl}. Hence, it was difficult to find molecules with an exact match. An interesting finding is that five (CHEMBL1583499, CHEMBL1726441, CHEMBL412355, CHEMBL2261013, and CHEMBL99068) of the 13 molecules had been reported as active for some targets. Among these, three (CHEMBL1583499, CHEMBL1726441, and CHEMBL412355) were active in the related roles in which the 5-HT\textsubscript{1B} docking molecules have been reported to have an effect. For example, CHEMBL1726441 is reported to be active for various targets such as Corticotropin-releasing factor receptor 2, Rap guanine nucleotide exchange factor 4, Nuclear factor erythroid 2-related factor 2, and Geminin. These targets have been reported to act in the human brain and peripheral tissues, playing a psychopathological role \cite{hillhouse2001control} and controlling brain function \cite{kawasaki1998family}. These investigations were conducted with the PubChem Bioassay database \cite{wang2012pubchem}. The other investigated results for the two remaining molecules (CHEMBL2261013 and CHEMBL99068) are summarized in Table S7\dag.

\subsection{Application to discovery of HIV inhibitors}\label{subresults6}

This section describes experiments in which RL-guided combinatorial chemistry was applied to discover HIV inhibitors. Here, we selected three HIV inhibition targets: C-C chemokine receptor type 5 (CCR5), HIV integrase (INT), and HIV reverse transcriptase (RT). Detailed information about the targets is summarized in Methods. To evaluate the HIV inhibition potency of molecules, pIC\textsubscript{50} predictors \cite{gottipati2020learning} for the three HIV inhibition targets were adopted. It should be noted that pIC\textsubscript{50} is equal to $-$logIC\textsubscript{50}, where IC\textsubscript{50} is an indicator that measures the amount of a particular inhibitory substance required to inhibit a given biological process or biological component by 50\%. In other words, the lower the value of IC\textsubscript{50}, the higher the HIV inhibition potency. Moreover, the higher the value of pIC\textsubscript{50}, the higher the HIV inhibition potency. Therefore, we set pIC\textsubscript{50} as the reward function to make our model discover HIV inhibitors with high potency.

In total, 10,000 generated molecules from our model were compared with the same number of molecules randomly combined by no-policy combinatorial chemistry (Fig. \ref{fig:HIV_results}a). For all the HIV inhibition targets, molecules generated by our model exhibited significantly higher pIC\textsubscript{50} values compared to those of random combination without a policy (original combinatorial chemistry). This result indicated that our model learned the fragment-selection policy to discover the intended substances. The benchmark results compared to the other four generators for HIV inhibitors are summarized in Table S9\dag, in which our model achieved the highest pIC\textsubscript{50} values for targets CCR5 and RT. The top 10 generated molecular structures with the lowest docking score are illustrated in Fig. S7\dag.

To analyze the policy change of molecular fragment selection in the training phase, we generated 10,000 molecules at the end of every training iteration. The derived frequencies (how many molecules had the fragment) for 25 fragments with the biggest change are plotted on the left of Fig. \ref{fig:HIV_results}b. In the initial state (where no policy was learned), the frequency of all fragments was similar. As the training progressed, the frequency of each fragment became varied. Although the selected frequency of some fragments increased as training proceeded, some of them decreased at certain points. This is because the agent found the combinations of molecular fragments that provided a high reward. Hence, the selection of other fragments that did not have any merit in the pIC\textsubscript{50} score rapidly decreased. The most selected fragments differed according to the type of HIV inhibition targets. Since the fragments were most often used to maximize the pIC\textsubscript{50} for each target, we hypothesize that they may be key structures for HIV inhibitors on each target.

\subsection{Expansion potential for discovery of organic materials
}\label{results_orgnanic_materials}

\begin{figure*}[!h]
    \centering
    \includegraphics[width=0.95\textwidth]{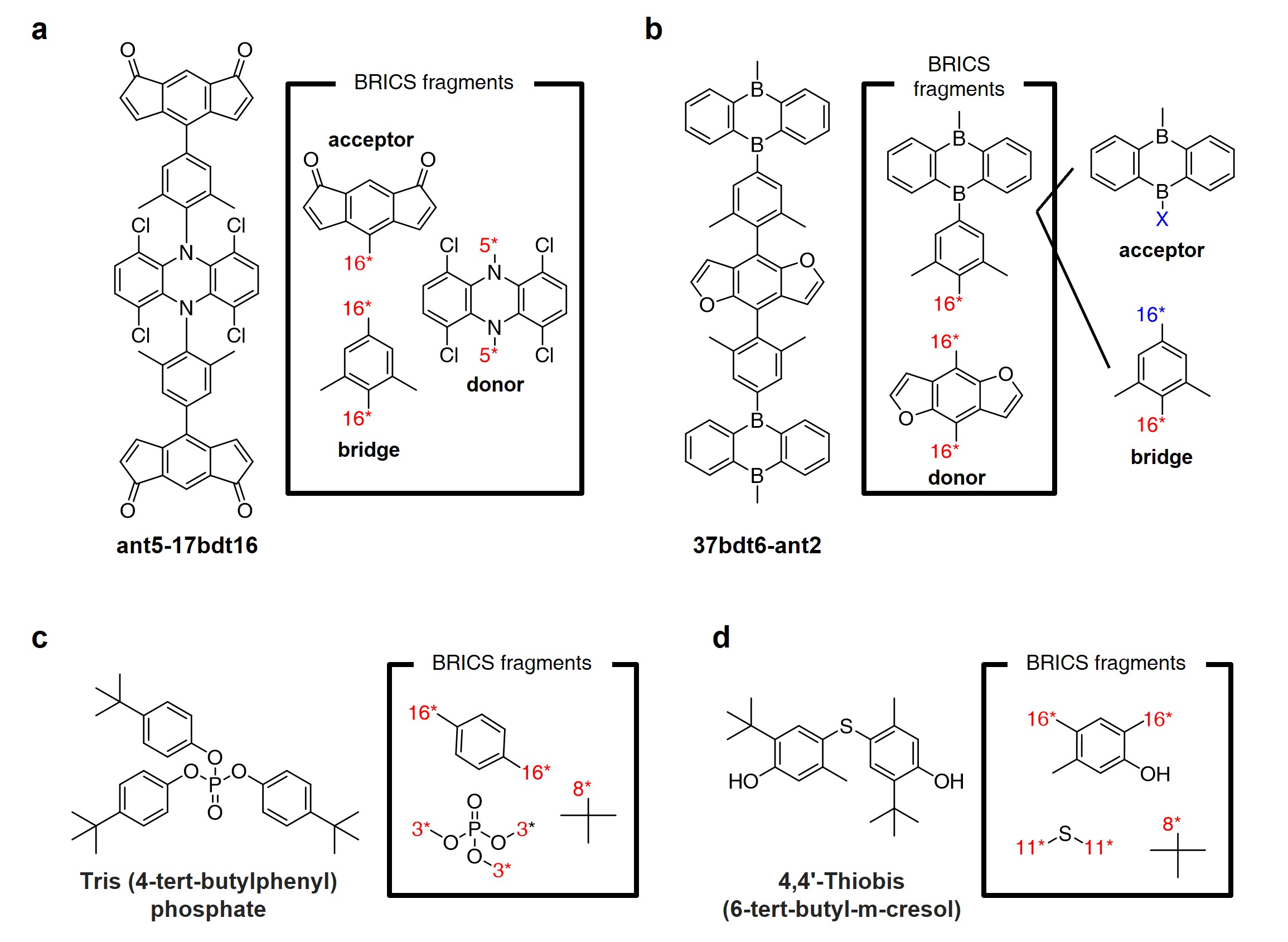}
    \caption{\textbf{BRICS fragmentation of organic materials.} (a,b) Thermally-activated delayed fluorescence (TADF) emitters for single-layer OLED reported in ref.\cite{tadf}. (c,d) Flame retardant materials reported in ref. \cite{flame_retardant}. 
    }\label{fig:organic_materials}
\end{figure*}

Fig. \ref{fig:organic_materials}a,b illustrate the results of fragmenting TADF emitters for single-layer OLED by leveraging the BRICS fragment combination rules in reverse. Examination of the fragments depicted in Fig. \ref{fig:organic_materials}a reveals that by inversely applying the BRICS fragment combination rules, it is possible to fragment the emitter into acceptor, donor, and bridge components, aligning with the knowledge of designing photoactive materials. However, the emitter shown in Fig. \ref{fig:organic_materials}b displays a chunk where the combination of acceptor and bridge remains undivided. This occurrence is due to the BRICS rules covering only the bonds between C, N, O, and S elements, lacking rules for bonds at the B-X binding sites marked in blue in Fig. \ref{fig:organic_materials}b. Consequently, this highlights a limitation that for the BRICS rules to encompass a wider range of organic materials beyond drugs, there is a need to incorporate additional rules for various elemental bonds. Fig. \ref{fig:organic_materials}c,d show the fragmentation results of organic flame retardant materials using the BRICS fragment combination rules in reverse. These results indicate that if the BRICS fragment combination rules are well-expanded and inversely applied to construct the fragment set of targeted materials, RL-CC has the potential to be sufficiently extended to the discovery of organic materials.

\section{Conclusion}\label{conclusion}
From a data science perspective, the discovery of better than previously known is to find new substances with properties outside outliers. In particular, research on inverse molecular design models with extrapolated target properties can be considered as fundamental groundwork for materials extrapolation. Most of the reported inverse molecular design models are based on data probability distribution-learning models, such as machine translators and generative models (including Seq2Seq \cite{sutskever2014seq2seq_original}, Transformer \cite{vaswani2017Transformer_original}, GAN \cite{goodfellow2014GAN_original}, and VAE \cite{kingma2013VAE_original}). However, these models are too limited for use in materials extrapolation, which requires discovering substances in the untrained area. To solve this problem, we adopted combinatorial chemistry \cite{combinatorial_chemistry}, which generates molecules from combinations of randomly selected molecular fragments. Fundamentally, it is a rule-based molecular designer for generating all chemically possible molecular structures that can be obtained from the combination of molecular fragments. However, since the lack of a molecular fragment selection policy can cause a combinatorial explosion \cite{klaus1986web}, RL is applied to train its fragment-selection policy to provide a direction toward target substances.

This paper contains three major contributions. First, we theoretically demonstrated that most inverse molecular design models based on probability distribution-learning of data are too limited for use in materials extrapolation. Second, we empirically demonstrated that our proposed RL-guided combinatorial chemistry works well on various discovery problems with extreme/extrapolated properties, such as the discovery of multiple target-hitting molecules, protein docking molecules, and HIV inhibitors. Since the BRICS \cite{degen2008BRICS} system is designed based on drug-like molecules, our applications were limited to the discovery of drug molecules. For these reasons, we have shown the feasibility of applying the proposed methodology to the discovery of organic materials by using the BRICS fragment bonding rules in reverse to decompose molecules of organic materials. This also highlights the limitations of the BRICS fragment combination rules that need to be addressed in order to further extend the scope of materials discovery. The problems addressed in this study exhibit significant overlap with practical materials discovery issues. For instance, photoactive materials used in applications such as organic light-emitting displays, solar cells, optical sensors, bioelectronic devices, and liquid crystal displays are compounds designed through combinations of molecular fragments. The common goals of these photoactive materials discovery are finding new and better materials that satisfy target properties, such as the energies of the S0, S1, T1 states, and their gaps. Additionally, considerations for band gaps in semiconductors and transport properties in battery electrolytes fall within these kinds of problems. Third, the limitations of our model were also analyzed in Methods. One of these was that re-training would be required to discover multiple target-hitting molecules if the targets were changed. The other limitation was a sparse reward problem that interrupted the discovery of materials with extreme properties. However, ongoing research efforts aimed at addressing these limitations suggest that they can be resolved through future studies.

\section{Methods}\label{methods}

\subsection{Molecular Descriptors}\label{Molecular Descriptors}

For the experiment in materials extrapolation, where the aim was to hit multiple extreme target properties, seven properties were set: logarithm of the calculated octanol-water partition coefficient (logP) \cite{logP}; topological polar surface area (TPSA) \cite{TPSA}; quantitative estimate of drug-likeness (QED) \cite{QED}; number of hydrogen bond acceptors (HBA); number of hydrogen bond donors (HBD); molecular weight (MW); and drug activity for dopamine receptor D2 (DRD2). Each property is considered important in the field of drug discovery. The term logP is a descriptor for the lipophilicity of a molecule, which refers to a molecule’s capacity to dissolve in fats or oils. This is an important property in drug design since it has an impact on the molecule’s capacity for penetrating cell membranes and reaching its intended target. The term TPSA is a calculated descriptor of the polar surface area (PSA) of a molecule, which refers to the area of a molecule having polar functional groups that could form hydrogen bonds with water molecules. A molecule is less polar and more likely to be able to penetrate cell membranes if it has a lower PSA value. The terms HBA and HBD are also important properties in drug design because they can affect a molecule’s capacity to interact with other molecules through hydrogen bonding. Hydrogen bonding is frequently used in drug design to facilitate the binding of a molecule to the target receptor. Moreover, hydrogen bonding can affect a drug molecule’s solubility and permeability, which influence its pharmacological properties. Term MW is a descriptor used in drug discovery, as it can affect a drug’s pharmacokinetics, efficacy, and safety. This is because the size of the molecules can influence a drug’s absorption, distribution, metabolism, or degree of penetration into the cell membrane. The correct molecular weight of a drug depends on its application. However, most drugs generally comprise small molecules with a molecular weight of less than 500 $Da$. This is because drug molecules have a higher likelihood of passing through a cell’s membrane and have a lower likelihood of being affected by biometabolic reactions. The term QED is a metric used to evaluate a molecule’s overall drug-likeness, which is a geomteric mean of logP, HBA, HBD, PSA, number of rotatable bonds (ROTB), number of aromatic rings (AROM), and number of structural alerts (ALERTS). A molecule with a high QED value is more likely to be a promising drug candidate. Finally, DRD2 refers to a drug's activity toward dopamine receptor D2. The dopaminergic neurotransmission is regulated by the G protein-coupled receptor dopamine receptor D2, which is mainly expressed in the brain.

In the experiment aimed at discovering protein docking molecules, QVina2 \cite{alhossary2015fast}\textemdash a tool to discover the minimum-energy docking conformation and calculate its docking score\textemdash was employed to compute a docking score. This score is proportional to the binding strength between a drug molecule and its target. QVina2 calculates the docking score by simulating how a drug molecule interacts with a given target receptor in a three-dimensional simulation box. We targeted the protein receptor 5-HT\textsubscript{1B}. Many studies have reported that activating 5-HT\textsubscript{1B} receptors outside the brain has vascular effects (such as pulmonary vasoconstriction, which can help treat angina \cite{morecroft19995}). Moreover, reduced 5-HT\textsubscript{1B} heteroreceptor activity can increase impulsive behavior, whereas reduced 5-HT\textsubscript{1B} autoreceptor activity can have an antidepressant-like effect \cite{nautiyal2015distinct, clark20015}.

In the experiment aimed at discovering HIV inhibitors, we selected three HIV-related targets: C-C chemokine receptor 5 (CCR5), HIV integrase (INT), and HIV reverse transcriptase (RT). Here, CCR5 is the immune system-related protein, which is found on the surface of white blood cells. Along with C-X-C chemokine receptor 4, it is a key co-receptor for HIV entry \cite{huang1996role}. The second target was INT, which is involved in viral replication and facilitates the viral cDNA’s insertion into the infected cells \cite{pommier2005integrase}. The final target was RT, which triggers the reverse transcription process. Here, the process can cause mutation and recombination that form the genetic diversity of HIV, enabling the formation of viral variants that could evade host immune responses, rendering the virus resistant to medication treatments \cite{sarafianos2009structure}. For each target in the experiment, we tried to maximize its pIC\textsubscript{50} value, which is a descriptor for the potency of a drug in inhibiting a biological activity. It is calculated as the negative logarithm of the IC\textsubscript{50} value, which is the amount of a drug that inhibits 50\% of the biological activity. In other words, the drug's potency increases as the IC\textsubscript{50} value decreases. In drug discovery, IC\textsubscript{50} is commonly utilized to compare the effectiveness of potential drug candidates.

\subsection{Fragment set configuration}\label{fragment set}
Two types of fragment sets were used in this study. One set contained 2,207 BRICS fragments that appeared more than 150 times in the training set \cite{kotsias2020cRNN} curated from the ChEMBL database \cite{mendez2019chembl} (release version 25). This fragment set was used for three major experiments: the discovery of seven target-hitting molecules, protein docking materials, and HIV inhibitors. For the supplementary experiment aimed at discovering five target-hitting molecules (ESI\dag 
~Note 1), another fragment set containing 2,547 BRICS fragments that appeared more than 100 times in the training set of the MOSES database \cite{polykovskiy2020moses} was used. The detailed reasoning for using these fragment sets is described in ESI\dag ~Note 2.

\subsection{Action masking}\label{action masking}
For efficient learning, we applied action masking to our model (stage 2 in Fig. \ref{fig:model_scheme}a). Based on the molecule in the current state, the list of actions that the agent can choose is limited by action masking. Since our model combines fragments according to BRICS fragment combination rules, the fragments that could not be connected to the molecule in the current state were masked. Thereby, more efficient learning was possible since the number of selectable actions in the current state was reduced, which improved the performance of our model (ESI\dag 
~Note 3).

\subsection{Target properties and calculation of molecular descriptors}\label{properties and calculation}

In this study, three major experiments were conducted to generate molecules with extreme properties. In the first experiment, discovering molecules that could hit the multiple target properties was attempted. In the other experiments, discoveries of molecules that maximize interesting properties were attempted. Accordingly, the interesting properties could be calculated by the evaluators in the environment of the RL model. The detailed information on model accuracy can be found in ESI\dag ~Note 7.

For the first experiment on materials extrapolation to hit multiple extreme target properties, seven molecular descriptors were set as the targets: logP \cite{logP}, TPSA \cite{TPSA}, QED \cite{QED}, HBA, HBD, MW, and DRD2 \cite{kotsias2020cRNN}. Here, the DRD2 was calculated using a QSAR model for DRD2 \cite{kotsias2020cRNN}, while the other descriptors were calculated using RDKit \cite{rdkit}. We selected 10 target sets of molecular properties out of known molecular data and attempted to generate the target-hitting molecules to demonstrate that our model would work well in terms of materials extrapolation. These 10 target sets of extreme properties were taken from an untrained dataset\textemdash PubChem SARS-CoV-2 clinical trials \cite{PubChemCovid19}\textemdash whose molecular properties are more widely distributed than the training data set \cite{kotsias2020cRNN} (Fig. \ref{fig:chembl_results}). Since the 10 target sets of extreme properties were taken from real molecules, they could be considered chemically feasible targets to discover.

For the experiment to discover protein docking materials, the calculated docking score between a docking molecule and 5-HT\textsubscript{1B} protein receptor was set as the target. The docking score was calculated using QVina2 \cite{alhossary2015fast}, which calculates the docking score of a docking molecule by searching for its minimum-energy docking conformation. This program employs an empirical scoring function to predict the docking score, which includes several terms that incorporate various physical and chemical interactions between the ligand and protein. These interactions include van der Waals interactions, electrostatic interactions, hydrogen bonds, and solvation effects. Moreover, QVina2 makes use of finely tuned scoring function parameters that are derived using many experimentally-identified ligand-protein complexes. It calculates a docking score for each docking posture it produces, with the lowest value being the most energetically favorable binding conformation. The detailed calculation configuration for QVina2 is illustrated in ESI\dag ~Note 4.

For the experiment to discover HIV inhibitors, the target property was set to maximize the pIC\textsubscript{50} score to three HIV-related targets: CCR5, INT, and RT. Each pIC\textsubscript{50} score was calculated by the light gradient boosting machine (LGBM) \cite{ke2017lightgbm}-based QSAR models \cite{yang2021hit} for the three HIV-related targets \cite{gottipati2020learning}. In addition, the QSAR model was trained to predict the pIC\textsubscript{50} value for each target using the ChEMBL dataset \cite{mendez2019chembl}.

\subsection{Training loop}\label{training loop}

To maximize the cumulative reward for sequential actions, RL trains its agent to learn which action to choose at each step. In this study, the action is a selection of a molecular fragment. The reward is the value calculated with the evaluator(s) by the experiment-specific reward function. By repeating this process, the policy that selects an action that can maximize the cumulative reward is gradually updated. After sufficient learning, the policy could then select sequential actions to generate a desired molecule that fits the given task.

The episode proceeds from steps 0 to $T$, where $T$ is the designated maximum fragment number of the molecule. To generate diverse molecules, the first fragment is randomly selected. When selecting the next molecular fragment to be combined with the current molecule, action masking is conducted. In the process of action masking, the molecular fragments that cannot be combined with the current molecule are masked according to the BRICS fragmentation combination rules. Subsequently, the policy selects a molecular fragment from the unmasked fragments and binds it to an unfilled binding site. If the combined molecule still has any unfilled binding sites, hydrogen atoms are attached to derive a complete molecule and evaluate its properties of the current state. When a molecule that meets the set of criteria is generated, the fragment attachment is stopped and the episode ends early before reaching step $T$. Otherwise, the process is repeated during step $T$. After the final output molecule is generated, a reward for the episode is calculated depending on how closely the targets are hit. Since the reward is given only at the end of an episode, the policy undergoes updates aimed at achieving a global solution rather than a local one. By iteratively conducting the training episodes, the agent of RL learns a policy that maximizes the given reward function. The episode training is repeated until a preset number of training iterations is reached. For each experiment, the training iteration was set to 750, 80, and 250 times.

Training for a predefined number of iterations was conducted using an Intel Xeon Gold 6226R. For each experiment, it took 12, 140, and 30 hours, respectively. The variation in training times arises from differences in the time required for property prediction and the batch size used at each training iteration. For each experiment, it takes an average of 0.35, 0.7, and 15 seconds, respectively, to predict the properties at each step. Additionally, the models’ accuracy can be found in ESI\dag~Note 7 and Table S10\dag.
For our model, retraining is required from scratch each time the target changes. Although transfer learning and other techniques for high efficiency may be considered in the future, they have not been applied as of now. In materials discovery, a fixed target is typically provided once the problem is given, enabling the training of the policy only once, and subsequent inference can be repeated at a low cost.

\subsection{Rewards and terminations}\label{rewards and terminations}
In RL, it is important to have an appropriate set of termination conditions and target conditions to learn a decent policy. In all three experiments conducted in this study, there were two common termination conditions. First, molecular generation was terminated when the number of fragments that make up a compound in the current state $n_{eval}$ exceeded the maximum number of fragments $n_{max}$, which was set to 50, 4, and 6 for each experiment. Second, the process was terminated when the number of unfilled binding sites $n_{L}$ was equal to zero.

For the experiment on materials extrapolation to hit multiple extreme target properties, the process was terminated if the $MW$ in the current state $MW_{eval}$ exceeded the maximum $MW$ ($MW_{max} = 3,500$ $Da$). Furthermore, it was also terminated if the error ($\epsilon$) was less than the error threshold $\epsilon_{th}=0.05$. Here, $\varepsilon$ was calculated as
\begin{equation}
\varepsilon = \sum_{y\in prop} \left ( \frac{y_{trg}-y_{eval}}{\sigma _{y}} \right )^{2}
\end{equation}
where $y_{trg}$, $y_{eval}$, and $ \sigma _{y}$ denote the target $y$, evaluated $y$, and standard deviation of $y$ for the curated ChEMBL training set \cite{kotsias2020cRNN}, respectively. Here, $prop$ is a set of properties that includes logP, TPSA, QED, HBA, HBD, MW, and DRD2.

The design of the reward function is also important since RL is performed based on the reward obtained by taking an action. Moreover, a penalty can be given to avoid any undesired actions. For the experiment on materials extrapolation to hit multiple extreme target properties, the reward function $r$ was designed as follows:
\begin{equation} \label{eqn:reward_1}
    r =\left\{\begin{matrix}
    &0,& &\textup{if} \left\{ \left [ MW_{eval} < MW_{min} \right ]\vee \left [ n_{eval} < n_{min} \right ]\right\}\wedge \left [ n_{L}\neq 0 \right ], \\
    &-50,& &\textup{if} \left\{ \left [ MW_{eval} < MW_{min} \right ]\vee \left [ n_{eval} < n_{min} \right ]\right\}\wedge \left [ n_{L}=  0 \right ], \\
    &\frac{100}{\varepsilon+1},& & \textup{else if} \ \varepsilon < \varepsilon_{th}, \\ 
    &\frac{30}{\varepsilon+1},& & \textup{otherwise.} \\
    \end{matrix}\right.
\end{equation}

Here, $MW_{min}$ and $n_{min}$ were set to generate various compounds by avoiding premature termination, which would generate uniform molecules that were too small. For this purpose, when the number of unfilled binding sites $n_L$ was not equal to zero, a zero reward was given if $MW_{eval}$ was less than $MW_{min}$ or $n_{eval}$ was less than $n_{min}$. When $n_L$ was equal to zero, a reward of $-50$ was given if $MW_{eval}$ was less than $MW_{min}$ or $n_{eval}$ was less than $n_{min}$. However, if a generated molecule did not correspond to the above two cases, a reward proportional to the degree of proximity to the target was awarded.

For the other two experiments (discovery of protein docking materials and HIV inhibitors), the reward function $r$ was designed as follows:
\begin{equation} \label{eqn:reward_2}
    r =\left\{\begin{matrix}
    &0,& &\textup{if} \left\{ \left [ MW_{eval} < MW_{min} \right ]\vee \left [ n_{eval} < n_{min} \right ]\right\}\wedge \left [ n_{L}\neq 0 \right ], \\
    &-50,& &\textup{if} \left\{ \left [ MW_{eval} < MW_{min} \right ]\vee \left [ n_{eval} < n_{min} \right ]\right\}\wedge \left [ n_{L}=  0 \right ], \\
    &\textup{pIC\textsubscript{50}},& & \textup{otherwise.} \\
    \end{matrix}\right.
\end{equation}

Here, the predicted score refers to the calculated docking and pIC\textsubscript{50} scores of the HIV-related target, respectively.

\subsection{RL algorithm} \label{algorithm}

We performed benchmark testing against the following state-of-the-art RL algorithms: IMPALA \cite{espeholt2018impala}, APPO \cite{schulman2017proximal}, A2C \& A3C \cite{mnih2016asynchronous}, and PPO \cite{schulman2017PPO}. The detailed results of the benchmarking are summarized in ESI\dag ~Note 3. From the benchmark results, we confirmed that PPO\textemdash which is a model-free, on-policy, actor-critic, and policy-gradient algorithm\textemdash was the most suitable for our problems, with a very large discrete action space of over 2,000. Moreover, PPO is known for its good performance, stability, and good sample efficiency, which makes the training process more stable by avoiding large policy updates with importance sampling and reusing learning data on the trust region. Hence, we applied PPO for all experiments conducted in this study. The objective function of PPO is defined as follows:

\begin{equation}\label{eqn:PPO}
\begin{split}
L^{CLIP}(\theta)=\hat{E}[\textup{min}(r_{t}(\theta )\hat{A_{t}}, clip(r_{t})(\theta ), 1-\epsilon , 1+\epsilon )\hat{A_{t}}], \\
\text{where}\ r_{t}(\theta)=\frac{\pi_{\theta}(a_{t}\mid s_{t})}{\pi_{\theta_{\textup{old}}}(a_{t}\mid s_{t})}
\end{split}
\end{equation}

\noindent where $\hat{E}_{t}$ represents the expected value at time step $t$. Term ${r}_{t}$ denotes the ratio between the new policy $\pi_{\theta}$ and the old policy $\pi_{\theta_{\textup{old}}}$. The policy is expressed as $\pi_{\theta }({a}_{t}\mid{s}_{t})$, where ${a}_{t}$ and ${s}_{t}$ are the action and state at timestep $t$, respectively. In Equation (\ref{eqn:PPO}), $\hat{A_{t}}$ denotes the advantage function at time step $t$, which estimates the result of the step action more effectively than the behavior of the default policy.

\subsection{Further findings}\label{further findings}

We empirically demonstrated that our methodology can discover novel molecules with extreme properties, which is impossible to accomplish with existing models that learn the probability distribution of data. However, there were two limitations with our model, which could be solved through further studies. First, the model should be re-trained if the target changes because the reward function depends on the given target. Therefore, the model must learn the policy from the start each time a new target is set. To solve this problem, a methodology such as meta-learning may be applied in future studies. With meta-learning, it is possible to predict the results of a new task or recommend hyperparameters based on the learning results from another task. Therefore, when a new task is assigned, it will be possible to learn through the experiences that have been taught in the past, requiring less additional training. For example, using methods such as MetaGenRL \cite{kirsch2019improving}, which applies meta-learning to reinforcement learning, the model outperformed the existing reinforcement learning algorithm while exhibiting similar performance for completely different tasks. Second, there was a sparse reward problem, which occurred because the agent received little feedback or reward for the action from the environment, rendering it difficult to learn efficient policies and satisfy the desired goal. Since molecules with extreme properties are rarer than common molecules, the probability of experiencing an episode in which a rare molecule is obtained through a random combination of molecular fragments is relatively low. To solve this problem, we could adopt methods that encourage more exploration with curiosity. This would allow experiencing more episodes that could provide higher rewards. In addition, hierarchical reinforcement learning \cite{riedmiller2018learning} could be applied, which is a methodology that utilizes prior knowledge of the given problem to set sub-goals that are easier to achieve than the original goal. Accordingly, it learns a policy that can achieve the original goal through policies learned from sub-goals.

\section*{Data availability}
The curated ChEMBL datasets and MOSES data sets for training and testing cRNN \cite{kotsias2020cRNN} and GCT \cite{Kim2021GCT} are publically available at \url{https://github.com/MolecularAI/Deep-Drug-Coder} and \url{https://github.com/molecularsets/moses}, respectively. All generated molecules in this study are available at \url{https://github.com/Haeyeon-Choi/RL-CC/tree/main/result}.

\section*{Code availability}
The full code used to perform the analysis is available at \url{https://github.com/Haeyeon-Choi/RL-CC}.

\section*{Author Contributions}
HK proposed the concept of materials extrapolation and the scheme of RL-guided combinatorial chemistry. HK, HC, and JN designed the experiments, and they were implemented by HK, HC, and DK. All authors analyzed the results and discussed them. HK, HC, and DK wrote the manuscript, and all authors reviewed it. JN and WBL supervised the project.

\section*{Conflicts of interest}
There are no conflicts to declare.

\section*{Acknowledgements}
This research was supported by the National Research Foundation of Korea (NRF) grant funded by the Korean Government through the Ministry of Science and ICT (MSIT) (NRF-2021R1C1C1012031, NRF-2021R1A4A3025742, NRF-2018M3D1A1058633, and NRF-2020M3F7A1094299).



\balance



\providecommand*{\mcitethebibliography}{\thebibliography}
\csname @ifundefined\endcsname{endmcitethebibliography}
{\let\endmcitethebibliography\endthebibliography}{}

\end{document}